\documentclass{aa}  

\usepackage{graphicx}
\usepackage{txfonts}
\usepackage{dsfont}
\usepackage{nicefrac}
\usepackage[english]{varioref}
\usepackage{hyperref}
\usepackage{booktabs}
\usepackage{relsize}
\usepackage{tikz}
\usepackage{pgfplots, mathtools}
\usetikzlibrary{arrows,decorations.pathmorphing,decorations.markings,backgrounds,positioning,fit,calc,shapes,shapes,matrix,pgfplots.groupplots,patterns}
\pgfplotsset{compat=newest}
\usepackage{subcaption}

\pgfplotscreateplotcyclelist{longcolorlist}{%
red,densely dashed,every mark/.append style={fill=red!80!black},mark=square*\\%
red,dashed,every mark/.append style={fill=red!80!black},mark=*\\%
red,every mark/.append style={fill=red!80!black},mark=star\\%
black,mark=diamond\\%
green,every mark/.append style={fill=green!80!black},mark=*\\%
blue,dashed,every mark/.append style={fill=blue!80!black},mark=*\\%
blue,every mark/.append style={fill=blue!80!black},mark=star\\%
}

\labelformat{chapter}{Chap.~#1}
\labelformat{section}{Sec.~#1}
\labelformat{appendix}{App.~#1}
\labelformat{subsection}{Sec.~#1}
\labelformat{subsubsection}{Sec.~#1}
\labelformat{figure}{Fig.~#1}
\labelformat{subfigure}{Fig.~\thefigure #1}
\labelformat{table}{Tab.~#1}
\labelformat{equation}{Eq.~(#1)}

\newcommand{\nifty}{\textsc{NIFTy}}
\newcommand{\imagine}{\textsc{Imagine}}
\newcommand{\hammurabi}{\textsc{Hammurabi}}
\newcommand{\hammurabiX}{\textsc{Hammurabi~X}}
\newcommand{\healpix}{\textsc{HEALPix}}
\newcommand{\multinest}{\textsc{MultiNest}}
\newcommand{\pymultinest}{\textsc{PyMultiNest}}
\newcommand{\python}{\textsc{Python}}

\newcommand{\tr}{\mathrm{tr}}

\begin{document}

   \title{Inferring Galactic magnetic field model parameters using IMAGINE}
   \subtitle{An Interstellar MAGnetic field INference Engine}
   
   \author{Theo Steininger \inst{1, 2}
   		   \and
   		   Torsten A. Enßlin \inst{1, 2}
		   \and 
           Maksim Greiner \inst{3}
           \and
		   Tess Jaffe \inst{4, 5}
		   \and
           Ellert van der Velden \inst{6, 7}
           \and
           Jiaxin Wang \inst{8, 9}
           \and
		   Marijke Haverkorn \inst{6}
    	   \and
		   Jörg R. Hörandel \inst{6, 10}
		   \and
		   Jens Jasche \inst{3}
		   \and
		   Jörg P. Rachen \inst{6}
           }

   \institute{Max Planck Institute for Astrophysics, 
              Karl-Schwarzschild-Str. 1, 
              85741 Garching, Germany
		  \and
		      Ludwig-Maximilians-Universität München, 
		      Geschwister-Scholl-Platz 1, 
		      80539, München, Germany
	      \and
    	      Excellence Cluster Universe, 
    	      Technische Universität München, 
    	      Boltzmannstrasse 2, 
    	      85748 Garching, Germany		      
          \and
          	  CRESST, NASA Goddard Space Flight Center,
        	  Greenbelt, MD 20771, USA
          \and
        	  Department of Astronomy, 
        	  University of Maryland, 
        	  College Park, MD 20742, USA        	  
          \and
          	  Department of Astrophysics/IMAPP, Radboud University,
              Heyendaalseweg 135,
              6525 AJ Nijmegen, The Netherlands
          \and
          	  Centre for Astrophysics and Supercomputing,
              Swinburne University of Technology,
              PO Box 218,
              Hawthorn, VIC 3122, Australia     
          \and     
              Scuola Internazionale Superiore di Studi Avanzati, 
              Via Bonomea 265, 
              34136 Trieste, Italy
          \and
	  	      Istituto Nazionale di Fisica Nucleare, Sezione di Trieste, 
              Via Bonomea 265, 
       		  34136 Trieste, Italy
       	  \and
         	  Nikhef, 
         	  Science Park Amsterdam, 
         	  1098 XG Amsterdam, The Netherlands	  
          }
                       
   \date{Received January 12, 2018; accepted 0000 00, 0000}

  \abstract
  {The Galactic magnetic field (GMF) has a huge impact on the evolution of the Milky Way. 
  Yet currently there exists no standard model for it, as its structure is not fully understood.
  In the past many parametric GMF models of varying complexity have been developed that all have been fitted to an individual set of observational data complicating comparability.}
  {Our goal is to systematize parameter inference of GMF models.
  We want to enable a statistical comparison of different models in the future, allow for simple refitting with respect to newly available data sets and thereby increase the research area's transparency. 
  We aim to make state-of-the-art Bayesian methods easily available and in particular to treat the statistics related to the random components of the GMF correctly.}
  {To achieve our goals, we built \imagine, the \emph{Interstellar Magnetic Field Inference Engine}.  
  It is a modular open source framework for doing inference on generic parametric models of the Galaxy. 
  We combine highly optimized tools and technology such as the \multinest\ sampler and the information field theory framework \nifty\ in order to leverage existing expertise.}
  {We demonstrate the steps needed for robust parameter inference and model comparison.
  Our results show how important the combination of complementary observables like synchrotron emission and Faraday depth is while building a model and fitting its parameters to data. 
  \imagine\ is open-source software available under the GNU General Public License v3 (GPL-3) at: \href{https://gitlab.mpcdf.mpg.de/ift/IMAGINE}{https://gitlab.mpcdf.mpg.de/ift/IMAGINE}}
   {}


   \maketitle
%

\section{Introduction}\label{sec:introduction}
The interstellar magnetic field in galaxies plays a key role in processes at various scales from star formation up to overall galactic evolution. 
Its energy density is comparable to that of the turbulent gas or cosmic rays (CRs), and therefore the dynamical feedback on the interstellar medium (ISM) must not be ignored. 
Galactic magnetic fields affect in- and outflows of the ISM that already exist as well as the formation of new ones.
They influence the propagation of CRs, which gyrate along the field lines. 
Though these effects are all important, it is challenging to infer the field, since it is only accessible via indirect detection methods. 
Additionally, since our Solar System is located within the Galactic plane, the tracers of the Galactic magnetic field (GMF) in our own Milky Way are highly degenerate as they are line-of-sight integrated quantities. 
This also means that the view of the opposite side of the Galaxy is obstructed by the intervening ISM. 
Because of all this, the GMF is currently mainly modeled via heuristic parametric models that have physically motivated features. 
The degrees of freedom in those models are morphological properties, field strengths (of possibly individual spatial components) of the magnetic field and the strength and characteristics of random contributions. 
Significant progress has been made here, which is the reason why a rather large number of GMF models is available today. 
At the same time the available data becomes better and better. 
Hence, there is need for a standardized platform that allows systematic parameter estimation and model comparison for a continuously expanding abundance of models and data.

\section{Bayesian Parameter Inference and Model Comparison}\label{sec:parameter_inference_model_comparison}
The GMF can naturally be thought of as a vector field with an infinite number of degrees of freedom:
    under the constraint of zero divergence the magnetic field can have an individual strength and direction at every point in space. 
This view corresponds to the most generic model possible, where the model's parameters are the field's degrees of freedom.
To infer the GMF one must simplify this most generic model, for example, by discretizing space. 
Doing so reduces the model parameters to a finite but still huge number, namely twice the number of voxels of the considered volume.
However, now one can try to concretely infer the magnetic field voxel by voxel, a method known as non-parametric modelling.
Generally speaking, constraining those non-parametric models is certainly hard, because the huge number of degrees of freedom often are counteracted by a limited amount of data. 
Because of this, one often builds a simpler model with a heavily reduced number of parameters, which therefore only covers a tiny slice in the full parameter space but still represents the most important features of the modeled quantity. 
In the case of the GMF, various models have been developed that differ greatly in their complexity: the number of parameters varies between only a few and up to 40.
Given a model and observational data one must find an estimate for a set of the model's parameters that explains the observed data well. 
However, in addition to the parameter estimation of a given model, there is also the task of comparing the plausibility of different models.
In the case of GMF inference this is especially important since so far there is no standard model available.

In terms of Bayesian inference, parameter estimation and model comparison can be described by the following components:
a given model $m$ that has a set of parameters $\theta$ shall be constrained by data $d$.
This means, that we are interested in the posterior probability density $P(\theta|d, m)$.
Bayes' theorem provides us with a calculation prescription 
\begin{equation}
    P(\theta|d, m) = \frac{P(d | \theta, m) P(\theta | m)}{P(d | m)},
\end{equation}
where $P(d | \theta, m)$ is the likelihood of the data, $P(\theta | m)$ is the parameter prior, and $P(d | m)$ is the model's evidence.
The latter guarantees the posterior's normalization and is given by
\begin{equation}
    \mathcal{Z} = P(d|m) = \int_{\Omega_\theta} P(d | \theta, m) P(\theta | m) \mathrm{d}\theta .
\end{equation}
For parameter estimation with one model, the evidence can be neglected, hence it is sufficient to maximize the product of the likelihood and the prior.
However, for comparing different models, e.g., $m_1$ and $m_2$, one needs normalized posteriors to form the ratio
\begin{equation}
    R = \frac{P(m_1 | d)}{P(m_2 | d)} = \frac{P(d | m_1) P(m_1)}{P(d | m_2) P(m_2)} = \frac{\mathcal{Z}_1 P(m_1)}{\mathcal{Z}_2 P(m_2)} . 
\end{equation}
Often there is no strong a priori reason for preferring one model over the other which corresponds to setting the model prior ratio $P(m_1)/P(m_2)$ to unity. 
In this case, the model's evidence is the only source of information for model selection. 

\section{Galactic Variance}\label{sec:galactic_variance}
The likelihood $P(d | \theta, m)$ describes the probability to measure the data $d$ if reality was given by $\theta$ and $m$. 
By modeling the physical system this probability can be explicitly calculated for certain sets ($\theta$, $m$). 
For this, one uses a forward simulation code to compute observables like sky-maps of Faraday rotation, synchrotron emission, and thermal dust emission.  
Given measured data, by modeling the noise characteristics of the detector, a probability can be assigned to the calculated maps, which is in principle a standard approach. 
However, when analyzing parametric models of the GMF one must be careful at this step because of how those models describe small scale structure of the magnetic field. 
Generally speaking, parametric models specify the large scale structure of the magnetic field explicitly by parameterizing the geometry of its components -- for example, the disk and possibly its arms, the halo, X-shaped components, et cetera -- and the field strength therein. 
Together, these components form the so-called regular field. 
Small scale structure, in contrast, is modeled in terms of its statistical properties rather than an explicit realization. 
This means, that when for a given parameter set $\theta$ a model instance is created, a random magnetic field is generated and added to the regular field. 
Depending on the model, the random magnetic field obeys, for example, a certain power spectrum, is locally proportional to the regular field, or shows a certain degree of anisotropy. 
As a consequence, the set ($\theta$, $m$) corresponds not only to one, but rather infinitely many possible field realizations. 
For the calculation of a likelihood this means that the measured observables must be compared with the ensemble average, which in practice is the simulated mean of a yet finite set of observable realizations that result from the magnetic field realizations. 
In theory one can work out the effect of various types of random fields on the used observables;
    for example, the total intensity of synchrotron emission does not depend on whether the structure of the magnetic field is ordered or completely random. 
Hence, one could use fudge factors to calculate the observable's mean directly without having to create numerous samples.
However, to do a proper uncertainty quantification one must not neglect the so-called \emph{Galactic variance},  a term introduced in \citet{jaffe:2010}. 
This variance measures how strong the influence of the random magnetic field on the individual pixels of an observable's sky-map is. 
Regions where the influence is high, that is where the observable's variance is high, must be down-weighted when being compared to measured data, in contrast to regions were the randomness of the magnetic field has little influence on the observable's randomness. 
This makes it again necessary to calculate instances of ($\theta$, $m$) to be able to construct an estimate for the Galactic variance. 
See \ref{sec:likelihood} for details.

\section{The \imagine\ Framework}

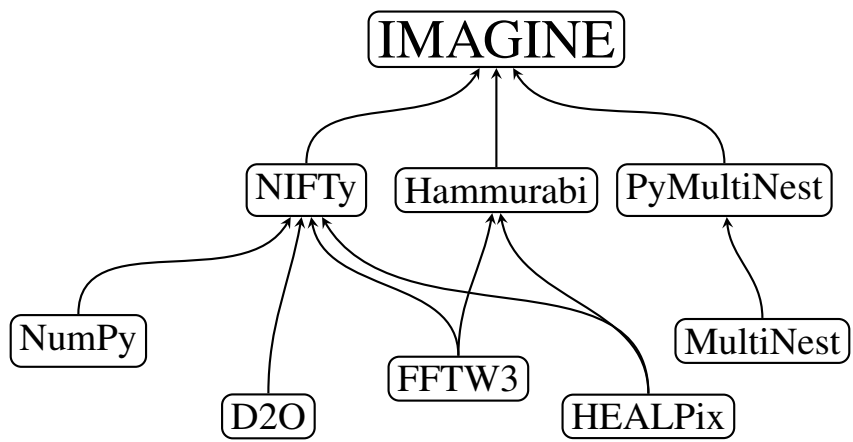
\begin{figure*}\centering
        \begin{tikzpicture}[>=stealth,thick,every node/.style={font=\relsize{2}, draw, shape=rectangle,rounded corners,align=center, anchor=center}]

        \node at (3, 2) (imagine) {\huge IMAGINE};

        \node at (0.5, 0) (nifty) {NIFTy};
        \node at (3, 0) (hammu) {Hammurabi};
        \node at (6, 0) (pymultinest) {PyMultiNest};

		\node at (6.5, -2) (multinest) {MultiNest};
        \node at (5, -3) (healpix) {HEALPix};
        \node at (-2.5, -2) (numpy) {NumPy};
        \node at (2.5, -2.5) (fftw) {FFTW3};
        \node at (0, -3) (d2o) {D2O};

        \draw[->] (healpix) to[out=90,in=-60] (nifty);
        \draw[->] (numpy) to[out=90,in=-120] (nifty);
        \draw[->] (fftw) to[out=90,in=-80] (nifty);
        \draw[->] (d2o) to[out=90,in=-100] (nifty);

        \draw[->] (fftw) to[out=90,in=-100] (hammu);
        \draw[->] (healpix) to[out=90,in=-80] (hammu);

        \draw[->] (multinest) to[out=90,in=-85] (pymultinest);

        \draw[->] (nifty) to[out=90,in=-120] (imagine);
        \draw[->] (hammu) to[out=90,in=-90] (imagine);
        \draw[->] (pymultinest) to[out=90,in=-60] (imagine);

        \end{tikzpicture}

      \caption{The building blocks of the \imagine\ framework.}
      \label{fig:imagine_building_blocks}

\end{figure*}

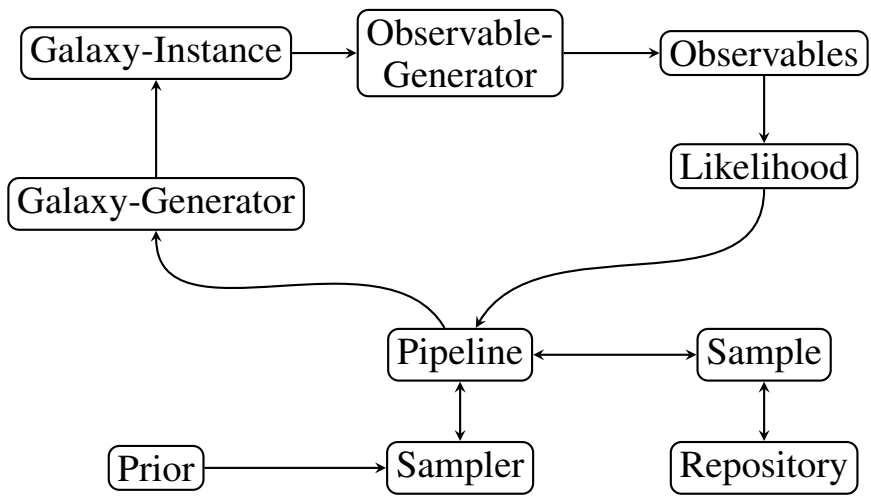
\begin{figure*}\centering
		\begin{tikzpicture}[>=stealth,thick,every node/.style={font=\relsize{2}, draw, shape=rectangle,rounded corners,align=center, anchor=center}]

		\node at (0, -1.5) (samp) {Sampler};
		\node at (-4, -1.5) (prior) {Prior};

		\node at (4, -1.5) (rep) {Repository};
		
		\node at (0, 0) (pipe) {Pipeline};
    	\node at (4, 0) (sample) {Sample};

		\node at (-4, 2) (bgen) {Galaxy-Generator};
		\node at (-4, 4) (bfield) {Galaxy-Instance};

		\node at (0, 4) (obsgen) {Observable-\\Generator};
		\node at (4, 4) (obs) {Observables};

		\node at (4, 2.5) (like) {Likelihood};

		\draw[->] (pipe) to[out=120,in=-90] (bgen);
		\draw[->] (bgen) to[out=90,in=-90] (bfield);
		\draw[->] (bfield) to[out=0,in=180] (obsgen);
		\draw[->] (obsgen) to[out=0,in=180] (obs);
		\draw[->] (obs) to[out=-90,in=90] (like);
		\draw[->] (like) to[out=-90,in=60] (pipe);
		\draw[<->] (pipe) to[out=0,in=180] (sample);
    	\draw[<->] (sample) to[out=-90,in=90] (rep);
		\draw[->] (prior) to[out=0,in=180] (samp);

		\draw[<->] (samp) to[out=90,in=-90] (pipe);
		
		\end{tikzpicture}

      \caption{The structure of the \imagine\ data processing and interpretation.}
      \label{fig:imagine_structure}

\end{figure*}

As mentioned in \ref{sec:introduction}, the number of available GMF models and the abundance and quality of observational data are continuously increasing. 
The goal of \imagine\ is to provide scientists with a standardized framework to analyze the probability distributions of model parameters based on physical observables. 
In doing so, Bayesian statistics is used to judge the mismatch between measured data and model prediction.
It is important to note that \imagine's inference is not limited to magnetic field models. 
Rather, \imagine\ creates an instance of the Milky Way based on a set of parameters.
It is irrelevant for the framework whether the parameters are controlling the appearance of the GMF or, for example, the properties of the free electron density or the dust density. 
Nevertheless, for the time being, we focus on the GMF and keep all other components fixed.  

It is desirable to have a flexible and open framework available when doing parameter inference. 
The magnetic field in particular must be analyzed indirectly via observables like synchrotron emission, Faraday rotation, dust absorption, or thermal dust emission since there is no direct detection method. 
This implies that the inference depends on the assumptions that were made regarding further constituents of the Milky Way, for example the free electron density, the population of cosmic rays, or the dust density. 
Hence, it is very likely that once the self-consistent analysis of a magnetic field model is finished, new insights regarding one or more other components make it necessary to redo the calculations with the new set-up. 
An example for this is the NE2001 model for the Galaxy's free electron density \citep{2004ASPC..317..211C}.
Today, updated versions like the YMW16 model \citep{2017ApJ...835...29Y} are available, and it would be very interesting to update parameter estimates from the past. 
In practice, either this does not happen at all or only with a huge time delay; inference pipelines are usually not made public and the originator may not have the necessary resources anymore. 
A standardized and open inference framework can help here to speed up scientific progress and make scientific results more transparent. 

\imagine\ is built on the programming language \python\ to ensure flexibility, and several external libraries for numerical efficiency, cf.\ \ref{fig:imagine_building_blocks}. 
Here, \python\ is primarily used as \emph{glue} to connect individual components and external libraries. 
A strictly object-oriented design makes it easy to extend its functionality from existing base-classes. 
The configuration of the inference runs is also done in \python.
No configuration files are used as the needs for future derived custom classes can not be foreseen today. 
Instead, the scientist instantiates the individual components in the main \python\ script which are ultimately embraced by the \imagine-pipeline.

\subsection{Components and Overall Structure}\label{sec:components_and_structure}
The structure of \imagine\ is shown in \ref{fig:imagine_structure} and discussed here.
The \texttt{Pipeline} object plays the key-role as it embraces all other objects and orchestrates their function calls. 
Its partner is the \texttt{Sampler}, with a functional interface for likelihood evaluations. 
The pipeline hides physical units and scales from the sampler. 
This means that the former exposes the latter $N$ variables ranging from $0$ to $1$, each.
In this way, the sampler can operate very generically on this unit cube $[0\dots 1]^N$ without the need to know any internal details on the Galaxy models. 

The likelihood evaluation inside the pipeline consists of the following steps. 
The \texttt{Sampler} yields a point from $[0\dots 1]^N$. 
Hence, first, the \texttt{Galaxy-Generator} maps those variables to physical parameters. 
Note that $N$ does not need to be the full number of all parameters a model has.
All parameters that are not marked as \emph{active} in the \texttt{Pipeline} are set to their individually configurable default value.
The \texttt{Galaxy-Generator} then uses these parameters to generate a certain \texttt{Galaxy} model realization. 
This means to set up all constituents of the abstract Galaxy model including,  
    e.g., the regular and the random magnetic field, the thermal electron density field, the dust-density field, et cetera. 
Next, the \texttt{Observable-Generator}, for example \hammurabi\ \citep{waelkens:2009}, processes the Galaxy instance and computes physical \texttt{Observables}, like sky-maps of the Faraday depth, synchrotron emission, or thermal dust emission. 
Those simulated quantities are then compared with measured data by the \texttt{Likelihood}, which in turn consists of sub-likelihoods for the individual observables. 
Together, parameters, Galaxy model, observables and likelihood values form a \texttt{Sample}.
Finally, the pipeline can be configured to store those \texttt{Samples} in a repository for post-processing and caching before the likelihood value is returned to the sampler. 
Together with the prior, the sampler can then determine which variable configuration should be evaluated next. 

As described in \ref{sec:galactic_variance} the GMF models may consist of a random field component to model the small scale structure stochastically. 
To deal with the resulting Galactic variance, instead of a single simulation, a set of realizations is created for a certain parameterization.
The members of that set are processed in parallel by the \texttt{Observable-Generator} such that horizontal scaling, i.e.\ using multiple computers as a cluster, can be exploited to compensate for the massively increased computational costs one has compared to approaches which ignore the Galactic variance. 
For this purpose, the \imagine\ framework uses the software packages \textsc{NIFTy\,3} \citep{2017arXiv170801073S} and \textsc{D2O} \citep{SGBE16} for convenient data processing and efficient data parallelization, respectively.  
\textsc{D2O} is based on the \emph{Message Passing Interface} standard (MPI) \citep{mpi-1-standard, mpi-2-standard} and in particular on mpi4py \citep{Dalcin20051108.mpi4py}.
In combination with OpenMP threading \citep{dagum1998openmp} of the \texttt{Observable-Generator} and the accompanying vertical scaling, \imagine\ efficiently exploits the parallel architecture of a modern high performance computing cluster as a whole as well as its nodes. 

\subsection{Using Sampling Methods for Uncertainty Quantification}
The goal of the \imagine\ framework is to provide deep probabilistic insights into Galaxy models given observational data. 
Because of the complexity of the problem, it is not sufficient to calculate point estimates like a maximum a-posteriori approximation.
We expect very counter intuitive interdependencies among the model parameters and hence need a thorough uncertainty quantification in order to correctly interpret the observations. 

To achieve this, \imagine\ uses Markov Chain Monte Carlo (MCMC) methods as described by \citet{gelman2014bayesian}.
As depicted in \ref{sec:parameter_inference_model_comparison}, we seek to perform parameter estimation for a given model as well as Bayesian hypothesis testing when comparing models. 
Because of its modularity, the \imagine\ framework can easily make use of the full arsenal of the Bayesian methodology, since it is straightforward to plug in different MCMC libraries and to write interfaces for new ones. 

Over the years, various sampling methods based on MCMC have been created.
In the following, we briefly discuss the concepts of \textit{Metropolis-Hastings}, \textit{Hamiltonian Monte Carlo} and \textit{Nested sampling}.

\subsubsection{Metropolis-Hasting Sampling}
The Metropolis-Hastings (MH) algorithm \citep{txt:Metropolis, txt:Hastings} creates a biased random walk through the parameter space. 
If the random walk is ergodic and its transition probabilities obey detailed balance, $P\left(\vec{x}\rightarrow\vec{x'}\right) P(\vec{x}) = P\left(\vec{x'}\rightarrow\vec{x}\right) P(\vec{x'})$, the samples generated by the random walk follow the probability distribution $P(\vec{x})$. 
Typically, this is achieved by combining a suggestion step with symmetric transition probabilities between any pair of locations from an unbiased random walk with a rejection step that ensures detailed balance, $P_{\mathrm{accept}} = \min\left\lbrace 1, P(\vec{x}_{\mathrm{proposed}})/P(\vec{x}_{\mathrm{old}})\right\rbrace$.

During the walk the samples in the chain must decorrelate from the starting position. 
Hence, the efficiency of an MCMC algorithm is crucial. 
Choosing a small step length for that purpose indeed means a lower rejection ratio. 
However, because of the small steps the chain does not move. 
In contrast, a large step length yields a high rejection ratio and therefore a chain that does not move, either. 
This relationship gets worse with higher dimensions. 
An approach to achieve high acceptance rates is Hamiltonian Monte Carlo sampling. 

\subsubsection{Hamiltonian Monte Carlo Sampling}
Hamiltonian Monte Carlo (HMC) sampling (also known as Hybrid Monte Carlo sampling) is a unique MCMC algorithm that introduces an auxiliary Gaussian random variable $\vec{p}$ of the same dimensionality as the original parameters $\vec{x}$, cf.\ \cite{brooks2011handbook, txt:Betancourt}.

The auxiliary variable plays the role of a momentum, the original parameters the role of a position in equations of motion from Hamiltonian mechanics. 
The negative log-probability corresponds to an energy. 
A new position in parameter space of position and momentum is generated by integrating the Hamiltonian equations of motion in time.
This new position is then treated as the result of a proposal step in the sense of the MH algorithm. 
Since the Newtonian equations of motions conserve energy the proposed parameters should be accepted $100\%$ of the time, while at the same time being far away from the initial parameters to ensure decorrelation of $\vec{x}$. 
This makes HMC sampling much more efficient in exploring the parameter space than MH sampling.

Although this makes an HMC sampler move much faster than an ordinary MH sampler it has a downside: it requires the gradient field of the desired probability density function (PDF).
Especially when dealing with a high number of dimensions, this can pose a problem if finite differencing must be used for gradient computation. 
Furthermore, some GMF models exhibit discontinuities that result in non-smooth likelihood landscapes, which makes gradients even more problematic. 
Hence, the \imagine\ pipeline primarily uses nested sampling which does not require gradient information and allows for model comparison, cf.\ \ref{sec:parameter_inference_model_comparison}, too. 

\subsubsection{Nested Sampling}
Nested sampling is an MCMC method developed by \citet{txt:Skilling}, that is capable of directly estimating the relation between the likelihood function and the prior mass.
It is unique in the fact that nested sampling is specifically made for usage in Bayesian problems, giving the evidence as its primary result instead of the posterior probability.

Nested sampling works with a set of \emph{live-points}. 
In each iteration, the point that has the lowest likelihood value gets replaced by a new one with a higher likelihood value. 
As this method progresses, the new points sample a smaller and smaller prior volume.
The algorithm thus traverses through nested shells of the likelihood. 

\subsection{Magnetic Field Models}
There are many parametric field models in the literature, from relatively simple axisymmetric spirals to complex multi-component models.
In addition to defining the parametrized structure of a magnetic field model, estimates for the values of those parameters must be made.
Usually, the term \emph{model} is used for both the analytical structure of the magnetic field and for a certain parameter fit. 
Note that in the context of \imagine, \emph{model} refers to the analytical structure only, since the goal is to investigate its parameter space.
It is more straightforward to denote two samples from the same parameter space as belonging to the same \emph{model} instead of constituting distinct models themselves, especially when doing Bayesian model comparison. 

In addition to the models' intrinsic complexities, the analyses in the literature also vary with respect to how many observables and datasets were used in the optimization.
An example for a rather simple magnetic field model that was fitted to only one observable is the WMAP logarithmic-spiral-arm (LSA) model \citep{page:2007}.
In a Galacto-centric cylindrical frame this regular GMF model is given as
\begin{align}
\vec{B}(r,\phi,z) &= B_0\left[\sin(\psi)\cos(\chi)\vec{\hat{r}} + \cos(\psi)\cos(\chi)\vec{\hat{\phi}} + \sin(\chi)\vec{\hat{z}}\right]~, \label{eq:WMAP}\\
\psi &= \psi_0 + \psi_1\ln\left(\frac{r}{R_0}\right)~,\nonumber\\
\chi &= \chi_0\tanh\left(\frac{z}{z_0}\right)~, \nonumber
\end{align}
where $\psi$ represents the pitch angle of the magnetic field spiral arm which varies according to $\psi_1$ and a logarithmic dependency on the radial distance $r$. 
 $R_0$ is the distance between the Galactic center and the Sun, and $\psi_0$ defines the local regular field orientation. 
The parameter $\chi$ corresponds to the off-disk tilting of the Galactic field, 
and $z_0$ characterizes the vertical scale height of the poloidal field strength modulation. 
This simple LSA model for the coherent field was fitted to synchrotron polarization data at $23~\mathrm{GHz}$ by \citet{page:2007}.  
Since the observable intensity of the synchrotron radiation depends on both $B_0$ and the cosmic ray electron (CRE) density in a degenerate way, only the other three parameters were fitted.

At the more complicated end is the \citet{jansson:2012c} model (JF12 hereafter) with dozens of parameters describing independent spiral arm segments for regular and random fields and thin and thick disks, an X-shaped halo, and more. 
JF12 was optimized against both Faraday rotation measures (RM) and synchrotron total and polarized intensity.  
The model of \citet{jaffe:2013} (and references therein, Jaffe13 hereafter) is in between in terms of number of parameters, with fewer fitted parameters compared to JF12 though originally optimized against the same observables.  

Some analyses in the literature include only a coherent field component, while some additionally study the random component from the turbulent ISM in a variety of ways.  
The JF12 model includes an analytic expression for the average amount of each observable that would result from the given turbulence model.  
Jaffe13 is notable in that it uniquely includes the effect of the Galactic variance described in \ref{sec:galactic_variance} explicitly in the likelihood.  That analysis used a set of numerical realizations of each model to quantify not only the average amount of emission but also its variations for a given point in parameter space, which is a necessary step for an unbiased likelihood analysis as described in \ref{sec:likelihood}.  

A further complication to this sort of analysis is how to treat the anisotropy in the random component. 
As described in \citet{jaffe:2010}, from an observational point of view, the GMF can be divided into three components: coherent, isotropic random, and a third variously called the ordered random, the anisotropic random, or the striated component.  
This third component is expected to arise in the turbulent ISM due to both shocks and shears on large scales.  
The JF12 model includes a scalar fudge-factor to adjust the synchrotron polarization amplitude from the coherent field to estimate this striated component.  
In contrast, Jaffe13 explicitly models it by projecting the numerically simulated isotropic random component onto the coherent component to generate an additional anisotropic component. 
These are complementary methods to model phenomenologically the effect of anisotropic, turbulent, magnetohydrodynamical processes that are computationally expensive to model physically.  

On an abstract level, the regular and random components of a magnetic field model are independent. 
Because of this, \imagine\ distinguishes them such that the user can combine any regular with any random field model. 
This is made possible not least through recent developments related to \imagine's primary observable generator \hammurabi. 

\subsection{Hammurabi}   
The \hammurabi\ code \citep{waelkens:2009} was built for simulating Galactic polarized foreground emission, absorption, and polarization rotation.
Its core functionality is to produce 2D observables in terms of \healpix\footnote{\href{ http://healpix.sourceforge.net}{http://healpix.sourceforge.net}} maps \citep{2005ApJ...622..759G} based on 3D physical field configurations in the Galaxy, e.g., the magnetic, cosmic ray and free electron fields.
To analyze various different models, \hammurabi\ is able to construct physical fields both analytically and numerically. 
Both regular and random fields covering Galactic scales can be generated with built-in field generators.
The observables are produced through line-of-sight integration, including synchrotron and polarized dust emission, Faraday depth, and dispersion measure.
In the course of the integration, radiative transfer and polarization rotation are evaluated by accumulating absorption and rotation effects backwards from the observer to the emitter.
Technically speaking, the line-of-sight integration is conducted on a set of nested \healpix\ shells. 
Given $R$ as the maximum simulation radius, the $n^{\mathrm{th}}$ shell out of $N$ total shells covers the radial distance from $2^{(n-N-1)}R$ to $2^{(n-N)}R$, except for the first shell which starts at the observer. 
The angular resolution in each shell is set by \healpix's $N_{\mathrm{side}}$ parameter. 
The $n^{\mathrm{th}}$ shell is by default set up with $N_{\mathrm{side}} = 2^{(n-1)}M$, where $M$ represents the lowest simulation resolution at the first shell. 
Accumulation of observables among shells is carried out by standard \healpix\ interpolation. 
Within each shell, physical quantities are estimated from inside out on discrete radial bins,
where the radial bin number is proportional to the radial thickness of the corresponding shell. 
Since the observables and the physical fields are constructed and evaluated in different coordinate frames,
a trilinear interpolation method is used to retrieve information from the physical fields during the line-of-sight integration.

\subsubsection{Random Magnetic Field Generation}\label{sec:hammurabi_random}
While exploring a magnetic field model's parameter space, the likelihood must be evaluated very often.
Hence, \hammurabi\ and especially its random field generator must be swift to preserve computational feasibility.  
To accomplish \imagine's scientific goals, \hammurabi\ was recently redesigned; the new version is called \hammurabiX\footnote{\href{https://bitbucket.org/hammurabicode/hamx}{https://bitbucket.org/hammurabicode/hamx}} hereafter.

In addition to numerous small to medium sized improvements, \hammurabiX\ provides two novel solutions for random magnetic field configurations on global, i.e.\ Galactic, and local, i.e.\ Solar neighborhood scales, respectively. 
In the case of global field generation, the focus lies on computational efficiency. 
Hence, a triple Fourier transform approach is used to do anisotropy enforcement, field strength rescaling and divergence cleaning.
For a given power spectrum, $P(k)$, a random magnetic field, $\vec{\tilde{B}}(\vec{k})$, is created in the harmonic Fourier base. 
The first Fourier transform translates $\vec{\tilde{B}}(\vec{k})$ into the spatial domain $\vec{B}(\vec{x})$.
There, anisotropy that may depend on the alignment of the regular magnetic field is introduced. 
Additionally, a template field strength scaling can be included in terms of a function $S(\mathbf{x})$ as
\begin{equation}
\vec{B}(\vec{x}) \rightarrow \vec{B}(\vec{x})\sqrt{S(\vec{x})} \text{.}
\end{equation}
An example for such a scaling function is 
\begin{equation}\label{eq:scaling_function}
S(\vec{x}) = S(r, \phi, z) = \exp\left(-\frac{r}{h_r} -\frac{|z|}{h_z}\right)\text{,}
\end{equation}
where $h_r$ and $h_z$ are the characteristic scales of the radial and vertical profiles, respectively. 
The second Fourier transform translates the re-profiled field $\vec{B(x)}$ back into harmonic space, where a Gram-Schmidt procedure is used to clean up the divergence:
\begin{equation}
\vec{\tilde{B}} \rightarrow \frac{\vec{\tilde{B}} - (\vec{k}\cdot\vec{\tilde{B}})\vec{k}/k^2}{\left|\vec{\tilde{B}} - (\vec{k}\cdot\vec{\tilde{B}})\vec{k}/k^2\right|} |\vec{\tilde{B}}| ~.
\end{equation}
Finally, a last Fourier transform is applied to retrieve the desired $\vec{B}(\vec{x})$.
Hence, the anisotropic random magnetic field is drawn from a one-dimensional power spectrum which in contrast corresponds to statistical homogeneity and isotropy. 
Breaking the isotropy with subsequent divergence cleaning results in a field that does not precisely obey the original power spectrum $P(k)$ anymore. 

In contrast to the global method, for local scale simulations a strict method including vector decomposition of the power spectrum tensor is available in \hammurabiX.
This method is not prone to the inaccuracies described above.
Details with respect to the local field generator are beyond the scope of this paper but are available in the release publication of \hammurabiX\ (Wang et al., in prep.). 

\subsection{Observables}
Magnetic fields cannot be measured directly.
Instead, their properties need to be inferred indirectly via \emph{observables} (also referred to as \emph{tracers}).
The most commonly used observables include Faraday rotation, synchrotron radiation, dust absorption and emission to probe properties of the GMF, as well as dispersion measure to probe the thermal electron density. 
These observables are briefly described below.

\subsubsection{Faraday Rotation}
Faraday rotation can be described as a double refraction effect when linearly polarized light travels through a magnetized, ionized medium.
The polarization angle of the Faraday rotation is given by
\begin{align}
\theta&=\theta_0+\Phi \lambda^2,
\end{align}
with $\theta$ being the observed polarization angle, $\theta_0$ the original polarization angle, $\Phi$ the Faraday depth and $\lambda$ the wavelength of the light ray.
The Faraday depth is given by a line-of-sight integral over a distance $l_0$ to an observer,
\begin{align}
\frac{\Phi}{\mathrm{rad\ m^{-2}}}&=0.812\int_{l_0}^0 \frac{n_e(l)}{\mathrm{cm^{-3}}}\frac{B_{\parallel}(l)}{\mathrm{\mu G}}\frac{\mathrm{d}l}{\mathrm{pc}},
\end{align}
with $n_e(l)$ and $B_{\parallel}(l)$ being the thermal electron density and strength of the parallel magnetic field, respectively, at distance $l$ away from the observer.
$\Phi$ is positive (negative) when the magnetic field is pointing towards (away from) the observer by convention.
Assuming the emitted polarization angle $\theta_0$ is constant for a specific source, the Faraday depth gives information about the average strength of the line-of-sight (i.e., parallel) component of the magnetic field.

\subsubsection{Synchrotron Radiation}
The synchrotron radiation that is used for the GMF inference is caused by the acceleration of relativistic electrons within this very magnetic field.
This linearly polarized electromagnetic radiation is emitted radially to the acceleration.
Its intensity is given by
\begin{equation} 
I_s \propto N\left(E\right)B_{\perp}^{x},
\end{equation}
with $N(E)$ being the density of relativistic electrons in the relevant energy range, $E$.
The index $x$ depends on the energy spectrum of these electrons, typically $x\approx 1.8$.
Even though the intensity of synchrotron radiation is degenerate with other emission components, like free-free and spinning dust in the microwave band, Stokes $Q$ and $U$ still provide information regarding the magnetic field. 
The other components are assumed to be unpolarized.
The random components of the GMF depolarize the synchrotron radiation; see the classic paper by \citet{burn:1966}.
The strength of this depolarization depends on the degree of ordering in the field, which can be written as $B_{\perp, r}^2/B_{\perp}^2$ with $B_{\perp, r}$ being the regular part of $B_{\perp}$.
Using the Stokes I, Q, and U together, we can calculate the strength of the magnetic field perpendicular to the line-of-sight $B_{\perp}$ (using the intensity $I$) and the fraction of the total magnetic field that is regular $B_{\perp, r}^2/B_{\perp}^2$ (using the polarized intensity $PI$).
This makes it a useful tool for studying the random component of magnetic fields.
In addition, the lines-of-sight for an extended source with a per se constant polarization angle traverse space with a different field configuration each. 
This results in varying polarization angles within the instrument beam, known as Faraday beam depolarization which provides further information.

\subsubsection{Dust Absorption and Emission}
Starlight polarization is caused by rotating dust grains absorbing certain polarizations of light.
In a magnetic field, a dust grain tends to align its long axis perpendicular to the direction of the local magnetic field (see \citet{txt:Davis} and references therein).
If the field is perpendicular to the line-of-sight, certain polarizations of the light-ray get blocked, viz.\ dust absorption of background starlight. 
The resulting observed light-ray is thus polarized, which gives information about the direction of the magnetic field perpendicular to the line-of-sight between the observer and the star.

The approach above works well for low-density dust clouds.
In high-density dust clouds, the probability that a light-ray gets completely absorbed along the way is fairly high.
However, dust heats up if it absorbs a lot of radiation, which in return will be re-emitted in the infrared. 
This emitted infrared light is also polarized according to the dust grain's geometry, viz.\ polarized thermal dust emission.
Since as already mentioned the dust grains are aligned in the magnetic field, the polarized dust emission provides complementary information about the direction of $B_{\perp}$.

\subsubsection{Dispersion Measure}
When a neutron star forms in the course of a supernova collapse the preserved angular momentum causes the neutron star to rotate rapidly. 
Along the neutron star's magnetic axis, a highly focused beam of radiation is emitted, and the rotational and magnetic axes are not necessarily the same. 
Since the beam is highly focused, from an observer's point of view this may result in a blinking pattern, which is why those stars are called pulsars. 
The group and phase velocity of the emitted radiation are not the same in the interstellar medium because of its ionized components, mainly free electrons. 
Because of this, higher frequencies arrive earlier than lower ones.
This extra time delay added at a frequency $\nu$ is given by
\begin{align}
t\left(\nu\right)&=\frac{e^2}{2\pi m_e c}\frac{\mathrm{DM}}{\nu^2},
\end{align}
with DM being the so-called dispersion measure.
The DM itself is given by the line-of-sight integral,
\begin{align}
\mathrm{DM}&=\int^{l_0}_0n_e(l)\ \mathrm{d}l.
\end{align}
If one has information on the thermal electron density, the DM solely depends on the distance $l_0$ between the source and the observer.

The dispersion measure, although it does not give any information on magnetic field properties, is still a very important observable.
With DM, the thermal electron density can be inferred, which in turn is needed for the inference of Faraday rotation, as described in \citet{1969ApJ...156L..21E}.
Using a combination of Faraday rotation, synchrotron radiation, starlight polarization and dispersion measure data is key for inferring the constituents of the Galaxy. 

\subsection{Likelihood}\label{sec:likelihood}
The likelihood is the probability $P(d | \theta, m)$ to obtain the data $d$ from a measurement under the assumption that reality is given by the model $m$ that in turn is configured by the parameters $\theta$.
It is the key element to rate the probability of a stochastic sample.
Assuming the generic case of a measurement with linear response function $R$ of a signal $s$ which involves additive noise $n$, the corresponding equation for the data $d$ reads 
\begin{equation}
	d = R(s) + n . 
\end{equation}
If the measurement device is assumed to exhibit Gaussian noise characteristics with a covariance matrix $N$, i.e.\
\begin{equation}
	n \hookleftarrow \mathcal{G}(n, N) = \frac{1}{\left|2\pi N \right|^{\nicefrac{1}{2}}} \exp \left(-\frac{1}{2}n^\dagger N^{-1} n\right)
\end{equation}
the log-likelihood for a simulated signal that is the result of the evaluation of a model $m$ with parameters $\theta$, i.e.\ $s' = m(\theta)$, to have produced the measured data $d$ is 
\begin{equation}\label{eq:simple_likelihood}
\mathcal{L}(d|s') = -\frac{1}{2}\left(d-R(s')\right)^\dagger N^{-1}\left(d-R(s')\right) - \frac{1}{2}\ln\left(\left|N\right|\right).
\end{equation}

In the context of \imagine, as discussed in \ref{sec:galactic_variance}, the GMF models posses random components that are described by ($m$, $\theta$) only stochastically. 
Marginalizing over those random degrees of freedom results in a modification of the effective covariance term in \ref{eq:simple_likelihood}, namely that the Galactic variance must be added to the data's noise covariance.
During the further discussion we consider the following quantities:
\begin{itemize}
\item The individual GMF samples within an ensemble of size $N_\mathrm{ens}$ are named $B^i$, with $i \in [1, N_\mathrm{ens}]$. 
\item The process of creating observables from $B^i$ is encoded in the response $R$. 
\item The simulated observables are denoted by $c^i = R(B^i)$. 
\item The measured observable's data is named $d$.
\end{itemize}

Denoting furthermore the data's noise covariance by $A$, the Galactic covariance by $C$, and the dimensionality of observables by $N_{\mathrm{dim}}$ the log-likelihood reads
\begin{equation}\label{eq:ensemble_likelihood}
	\mathcal{L}(d|{c}) = -\frac{1}{2} (d-\bar{c})^\dagger (A+C)^{-1} (d-\bar{c}) - \frac{1}{2}\ln\left(\left|A+C\right|\right)
\end{equation}
with the ensemble mean of $c$
\begin{equation}
\bar{c} = \frac{1}{N_{\mathrm{ens}}}\sum^{N_{\mathrm{ens}}}_{i=1} c^i \mathrm{.} 
\end{equation}
As discussed in \ref{sec:galactic_variance} the Galactic covariance $C$ reflects the fact that the observables posses an intrinsic variance because of the random parts of the GMF. 
For example, the higher the intrinsic variance, the more the likelihood will be flattened by the $(A+C)^{-1}$ term.
This means that the likelihood is less responsive to deviations from the ensemble mean for regions of high variance. 
Hence, there is the risk of overestimating random field contributions, since they are favored by the likelihood.
However, this is compensated by the second summand in \ref{eq:ensemble_likelihood}: the covariance matrix' log-determinant $\ln\left(\left|A+C\right|\right)$. 
In \ref{eq:simple_likelihood} the covariance matrix and thus its determinant are constant and therefore can be neglected as we are not interested in the absolute scales of the likelihood. 
In contrast, for \ref{eq:ensemble_likelihood} we have to consider it as this determinant varies from point to point in parameter space. 

The Galactic covariance $C$ is not known, hence, we must estimate it. 
A classic approach for $C$ is to evaluate the dyadic product of the samples' deviations from their mean:
\begin{equation}
\displaystyle C_{\mathrm{cl}} = \frac{N_{\mathrm{dim}}}{N_{\mathrm{ens}}} \sum^{N_{\mathrm{ens}}}_{i=1}(c^i - \bar{c})(c^i - \bar{c})^{\dagger} = \frac{1}{N_{\mathrm{ens}}}\sum^{N_{\mathrm{ens}}}_{i=1} u^i {u^i}^\dagger
\end{equation}
with
\begin{equation}
\quad u^i = \sqrt{N_{\mathrm{dim}}} \left(c^i - \bar{c}\right)\text{.}
\end{equation}	
Since the number of samples in an ensemble is much smaller than the number of dimensions this classical estimator for the covariance matrix is insufficient. 
Most of its eigenvalues are zero, making an operator-inversion impossible. 
Hence, it is better to use a sophisticated estimator using a shrinkage target (e.g., a diagonal matrix) and a shrinkage factor.
Here, we use the \emph{Oracle Approximating Shrinkage} (OAS) estimator by \citet{shrinkage}:
\begin{equation}
	C = \mu\rho~\mathds{1} + (1-\rho)~C_{\mathrm{cl}} \text{.}
\end{equation}
The specific quantities needed to compute the OAS estimator are
\begin{align}
\mu &= \frac{1}{N_{\mathrm{dim}}} \tr\left(C_{\mathrm{cl}}\right) = \frac{1}{N_{\mathrm{dim}} N_{\mathrm{ens}}}\sum^{N_{\mathrm{ens}}}_{i=1} {u^i}^\dagger u^i\\
a &= \tr\left(C_{\mathrm{cl}}^\dagger C_{\mathrm{cl}}\right) = \frac{1}{N_{\mathrm{ens}}^2} \sum^{N_{\mathrm{ens}}}_{i=1}\sum^{N_{\mathrm{ens}}}_{j=1} \left({u^i}^\dagger u^j\right)^2\\
r &= \min\left\lbrace1, \frac{\left(1-2/N_{\mathrm{dim}}\right)a + N^2_{\mathrm{dim}}\mu^2 }{\left(N_{\mathrm{ens}}+1-2/N_{\mathrm{dim}}\right)\left(a-N_{\mathrm{dim}}\mu^2\right)}\right\rbrace \text{.}
\end{align}
In the likelihood one needs to apply the inverse of the sum of $A$ and $C$, $(A+C)^{-1}$.
Since we do not know a basis in which $A+C$ is diagonal, the inversion of this operator is a nontrivial task.
However, because of its structure, we can use the Sherman-Morrison-Woodbury matrix identity \citep{sherman1950, woodburymaxa.1950} by re-sorting
\begin{equation}
	A + C = (A + \mu r~\mathds{1}) + (1-r) C_{\mathrm{cl}} = B + V V^\dagger
\end{equation}
with 
\begin{equation}
	B = A + \mu r~\mathds{1} \quad \text{and} \quad V = \sqrt{\frac{1-r}{N_{\mathrm{ens}}}} ~ U \text{.}
\end{equation}
Namely,
\begin{equation}\label{eq:covariance_full_inverse}
	(B + VV^\dagger)^{-1} = B^{-1} - B^{-1} V (\mathds{1} + V^\dagger B^{-1} V)^{-1} V^\dagger B^{-1} \text{.}
\end{equation}
With this formula only a matrix of size $N_{\mathrm{ens}}^2$ instead of $N_{\mathrm{dim}}^2$ must be inverted. 

For computing the log-determinant $\ln\left(\left|A+C\right|\right)$ one could use the result of the OAS estimator and apply the generalized form of the matrix determinant lemma \citep{harville2008matrix} to it. 
Its structure is closely related to the Sherman-Morrison-Woodburry matrix identity: it turns the problem into the calculation of the determinant of a matrix of size $N_{\mathrm{ens}}^2$ instead of $N_{\mathrm{dim}}^2$. 
For our case it reads:
\begin{equation}\label{eq:determinant_lemma}
	\left| A + C \right| = \left| B + VV^\dagger \right| = \left| B \right| \cdot \left| \mathds{1} + V^\dagger B^{-1} V \right|
\end{equation}
However, the OAS estimator has been designed for and is good at approximating covariance matrices in terms of quadratic forms; using it for determinant estimation yields rather poor results. 
And in fact, it can be shown that it is not possible to construct a general purpose estimator from covariance matrix samples if the number of samples is lower than the number of dimensions \citep{CAI2015161}. 
Nevertheless, heuristic as well as Bayesian estimators have been developed trying to cover special cases, as for example the case of sparse or diagonally dominated covariance matrices \citep{2017arXiv170401445F, hu2017comparison}. 
For the time being we approximate the determinant $\left| A + C \right|$ by its diagonal:
\begin{equation}\label{eq:diagonal_determinant}
\ln\left(\left| A + C \right|\right) \approx \frac{1}{N_{\mathrm{dim}}} \tr\left[\ln\left(A + \frac{1}{N_{\mathrm{ens}}}\sum_{i=1}^{N_{\mathrm{ens}}} \left(c^i-\bar{c}\right)^2\right)\right] \text{.}
\end{equation}
This approximation serves the purpose of regularizing the random magnetic field strength. 
Future improvements could include the usage of one of the widely used shrinkage estimators as discussed in \citet{hu2017comparison}. 
They work similarly to the OAS estimator, though exhibiting shrinkage coefficients and targets tailor made for covariance determinant approximation. 
For those, then \ref{eq:determinant_lemma} can be used for efficient computation. 
In either case, the inversion of the covariance matrix as well as the calculation of its determinant can be done explicitly, if approximately, which therefore allows us to evaluate the ensemble likelihood in \ref{eq:ensemble_likelihood} efficiently. 

\section{Application}\label{sec:application}
In the following we discuss possible usage scenarios of the \imagine\ pipeline.
Regardless of parameter estimation or model comparison, first, a Galaxy model must be set up. 
Below we will use \imagine\ to analyze the following scenario.  
Our Galaxy model consists of the WMAP logarithmic-spiral-arm (LSA) magnetic field model \citep{page:2007} in combination with an isotropic Gaussian random field as described in \ref{sec:hammurabi_random}. 
In \hammurabiX, the random field's normalization is chosen such that its RMS field strength at the Sun's position is given by $\tau$.  
We denote the spectral index of the random field's power spectrum as $\alpha$. 
Furthermore, we choose the YMW16 model \citep{2017ApJ...835...29Y} for the thermal electron density. 
Here, our goal is to infer the parameters of the magnetic field model, so the thermal electron density we assume to be fixed. 
For the input data, we consider polarized synchrotron emission at $1.41~\mathrm{GHz}$ (Stokes $Q$ and $U$) following \citet{2006A&A...448..411W}, $408~\mathrm{MHz}$ (Stokes $I$) and at $30~\mathrm{GHz}$ (Stokes $Q$ and $U$) following \citep{2016A&A...594A...1P}, and the Faraday depth map following \citet{2012A&A...542A..93O}. 

\subsection{Mock Data Based Tests}\label{sec:mock_data_tests}
\begin{figure*}
\caption{Simulated synchrotron emission difference maps (including $408~\mathrm{MHz}$ total intensity $T_{\mathrm{tot}}$ at northern hemisphere, $30~\mathrm{GHz}$ Stokes $Q$ and polarized intensity $T_{\mathrm{pol}}$ in $\mathrm{mK}$) with different $\psi_0$ or $\chi_0$ settings. $\psi_0$ has influence mainly along Galactic longitude while $\chi_0$ affects more the latitude direction.}
\resizebox{0.5\hsize}{!}{\includegraphics{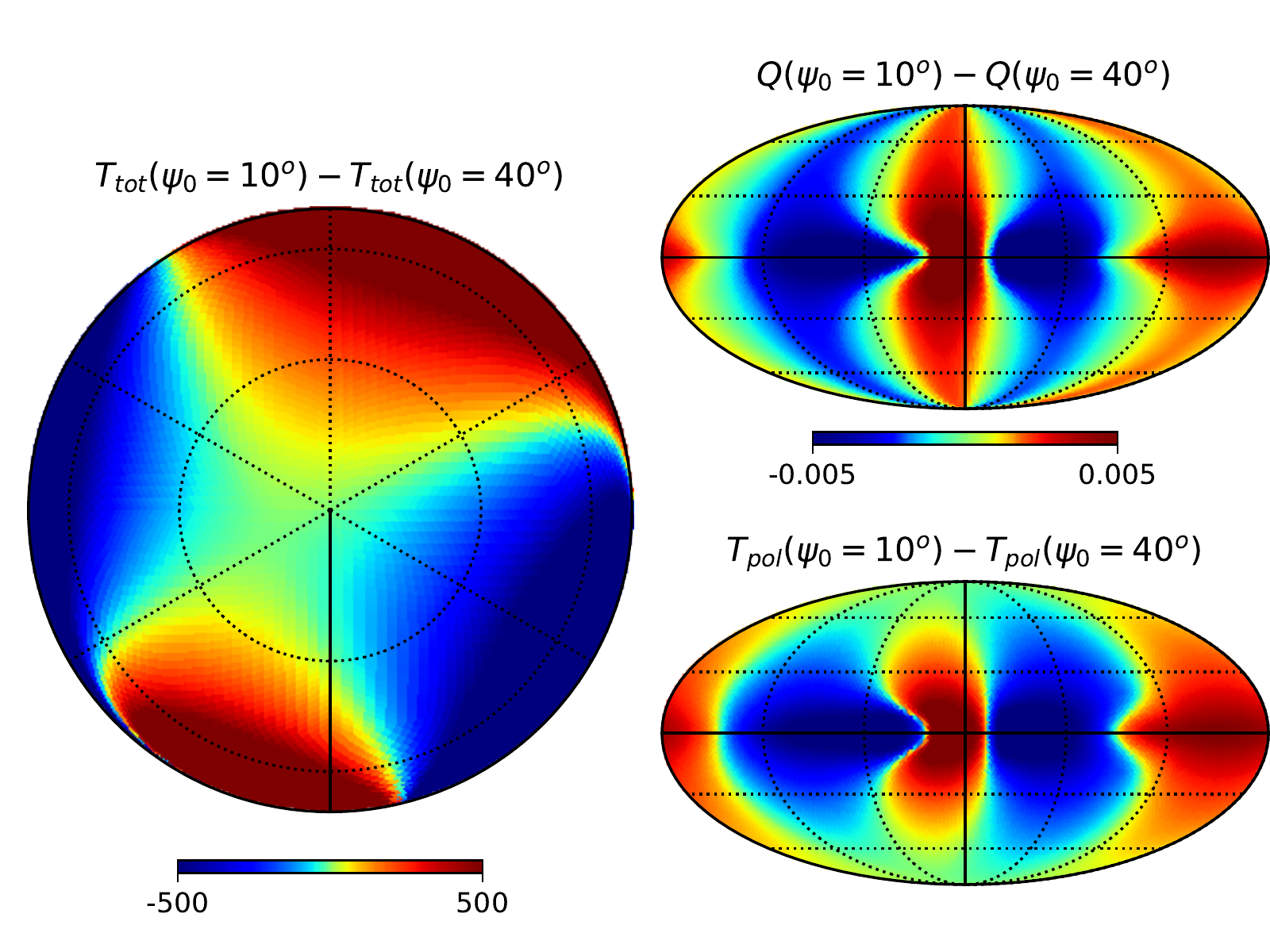}}
\resizebox{0.5\hsize}{!}{\includegraphics{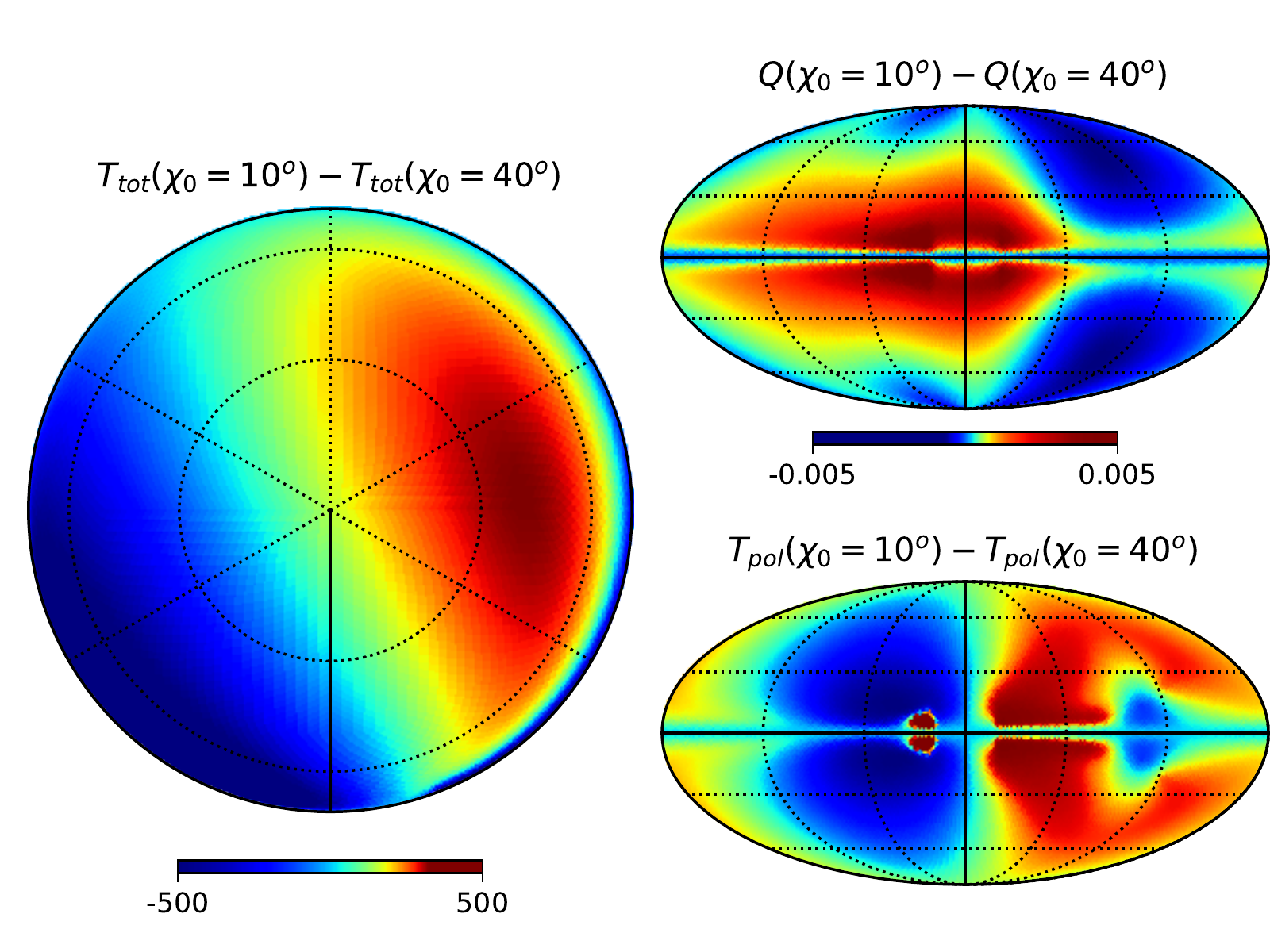}}
\label{fig:wmap_lsa}
\end{figure*}

It is advisable, before starting a large likelihood exploration, to check if the chosen observables (tracers) are sensitive to the model parameters that are about to be inferred. 
In principle, all observables used here are sensitive to the GMF configuration especially near the Solar neighborhood. 
In terms of the WMAP LSA model, the influence of $\psi_0$ on polarized synchrotron emission is expected to be the most noticeable feature, cf.\ \ref{fig:wmap_lsa}.
By definition of the model, $\psi_1$ has greater influence than $\psi_0$ and $\chi_0$ on the field's configuration when $r<R_0/e$. 
We therefore expect the observables to be more sensitive to $\psi_1$ at low Galactic latitudes where line-of-sight integration accumulates information through the Galactic center.  
However, Faraday depolarization at low Galactic latitudes and low frequencies diminishes constraining power of polarized synchrotron emission on $\psi_1$.

\subsubsection{Mock Data Configuration}
During the development of the IMAGINE framework, initial mock data tests were performed on the base of the JF12 model and are described in \citet{velden2017thesis}. 
One result of that work was an increased appreciation for the difficulty working with such a complex model. 
The first ($\omega_1$) is solely a regular WMAP field, while the second ($\omega_2$) additionally possesses a random component as described in \ref{sec:hammurabi_random}. 
To test the pipeline with a parameter set that is as realistic as possible, we used the best fit estimates for the WMAP LSA model given in \citet{page:2007}, except for $\psi_1$ which would be $0.9^\circ$. 
To conduct proper tests it is helpful if the mock data generating parameter values are not located at the boundaries of parameter space, so we set $\psi_1$ to $7.95^\circ$.
Since $B_0$ is not given in \citet{page:2007} we use \citet{Beck:2005db} as a reference and set it to $6~\mathrm{\mu G}$. 
Furthermore, for $\omega_2$ we set the random magnetic field's strength around the Sun $\tau$ to $2~\mathrm{\mu G}$.
The spectral index is set to $\alpha = 1.7 \approx 5/3$ (Kolmogorov).
The precise mock data parameter values and the boundaries of the tested parameter volume are given as follows
\begin{eqnarray}
	B_0 &=& 6.00\in [0.3, 11.7]~\mathrm{\mu G}\\
	\chi_0 &=& 25.0 \in [1.0, 49.0]^\circ\\
	\psi_0 &=& 27.0 \in [6.0, 48.0]^\circ\\
	\psi_1 &=& 7.95 \in [0, 15.9]^\circ\\
	\tau &=& 2.00 \in [0.2, 3.8]~\mathrm{\mu G}\\
	\alpha &=& 1.7 \in [0.2, 3.2]	
\end{eqnarray}

After processing the mock magnetic fields with \hammurabiX, we add individual random noise samples with the variances given in \citet{2006A&A...448..411W, 2016A&A...594A...1P, 2012A&A...542A..93O} to the calculated observables. 
For the Oppermann Faraday depth map there is an uncertainty map available which is based on a Bayesian Wiener filter reconstruction. 
Since the pixel-wise noise is uncorrelated on small scales, we downscale the uncertainty map to $N_\mathrm{side}=32$ to estimate the total noise power correctly. 
Since we produce the sample simulations with $N_\mathrm{side}=32$ as well, no further adaption of this noise map is necessary. 
For the Planck and Wolleben synchrotron (Stokes $Q$ and $U$ in each case) we take a constant statistical uncertainty of $2.12~\mu \mathrm{K}$ \citep[Tab. 10]{2016A&A...594A..10P} and $12~\mathrm{mK}$ \citep[section 5.2]{2006A&A...448..411W}. 
For the Stokes $I$ map at $408~\mathrm{MHz}$ an uncertainty map is given. 
We downgrade all four data sets to our simulation resolution of $N_\mathrm{side}=32$. 

For the inference below, the ensemble size was set to $N_{\mathrm{ens}} = 64$. 
Our tests showed that for the resolution $N_\mathrm{side}=32$ this is the ensemble size where the classical covariance term in the ensemble likelihood becomes dominant over the shrinkage target, i.e.\ $r$ falls below $0.5$, cf.\ \ref{sec:likelihood}.  
To make likelihood maximization and sampling possible, it is also necessary to stabilize the likelihood by fixing the ensemble member's random seed. 
This introduces a bias, which we found, however, to be already negligible in the case of $N_{\mathrm{ens}} = 64$ compared to the emerging Galactic variance. 
In the future, one could try to enhance existing sampling techniques already including simulated annealing \citep{1983Sci...220..671K} to become capable of treating the noisy likelihood surface directly. 

\subsubsection{Regular Magnetic Field}\label{sec:regular_mock_tests}
First, we consider the first mock data set $\omega_1$ that does not contain random field components. 
For this data set we perform one-dimensional likelihood scans through the parameter space, as this is a systematic way to check the observables' sensitivity with respect to the model parameters. 
In doing so, we vary only one parameter at a time while keeping all others fixed to the mock data's generating values. 

\ref{fig:parameter_scan_norandom} shows how well the different observables yield peaks in the likelihood.
Since there is no random magnetic field, the ensemble likelihood simplifies to a standard $\chi^2$ likelihood. 
Several comments are in order.
First, one sees that the total log-likelihood exhibits clear peaks very near to the \emph{true} mock data values for all four WMAP LSA parameters. 
Second, as expected, $B_0$ shows the strongest dependence, followed by $\psi_0$ and $\chi_0$; $\psi_1$ affects the observables as well but much more weakly than the other three parameters. 
Third, it is remarkable that for all parameters the total log-likelihood is dominated by synchrotron emission Stokes $Q$ \& $U$ at $30~\mathrm{GHz}$ and Stokes $I$ at $408~\mathrm{MHz}$. 
Faraday rotation also adds some information, but synchrotron data at $1.41~\mathrm{GHz}$ yields four to six orders of magnitude weaker signals in the log-likelihood. 
This is because of the signal-to-noise ration which is better for the Planck than for the Wolleben data set.
Furthermore, due to the depolarization effects that have a huge impact on low-frequency polarized synchrotron data, we see sharp peaks for $1.41~\mathrm{GHz}$ synchrotron data, as the morphology of the observable map tremendously changes when varying the magnetic field. 
If the GMF were regular, this would allow us to constrain the GMF parameters very precisely. 
However, the presence of random magnetic fields and Faraday depolarization effects render this frequency uninformative for this analysis. 
We therefore exclude the $1.41~\mathrm{GHz}$ data from the subsequent analysis.

\begin{figure*}[t]
\caption{\textbf{Mock data $\omega_1$:} Scans through the parameter space of the WMAP LSA regular magnetic field model, including simulated additive measurement noise according to \citet{2006A&A...448..411W}, \citep{2016A&A...594A...1P}, \citet{2012A&A...542A..93O}. 
The mock data input parameter values are at the very center of each abscissa and indicated by the vertical line. 
Since the Planck synchrotron data dominates the overall likelihood, we show the total likelihood in the leftmost plot.
Note that the log-likelihood varies over several orders of magnitude.}

	\begin{tikzpicture}
	\begin{groupplot}[group style = {group size = 3 by 4, horizontal sep = 30pt, vertical sep = 38pt}, width = 0.355\linewidth, table/search path={data/parameter_scan_norandom/}, cycle list name=longcolorlist]
		\nextgroupplot[xlabel={$B_0$ [$\mathrm{\mu G}$]}, ylabel={Log-Likelihood},ymin=-4e8]
			\addplot table [x={b0_sync_I_0.408_x}, y={b0_sync_I_0.408_y}] {data.dat};\label{plots:scan_norandom_1}
			\addplot table [x={b0_sync_Q_30_x}, y={b0_sync_Q_30_y}] {data.dat};\label{plots:scan_norandom_2}
			\addplot table [x={b0_sync_U_30_x}, y={b0_sync_U_30_y}] {data.dat};\label{plots:scan_norandom_3}
			\addplot table [x={b0_total_x}, y={b0_total_y}] {data.dat};\label{plots:scan_norandom_4}
   	        \draw (6, \pgfkeysvalueof{/pgfplots/ymin}) -- (6, \pgfkeysvalueof{/pgfplots/ymax});
		\coordinate (top) at (rel axis cs:0,1);

		\nextgroupplot[xlabel={$B_0$ [$\mathrm{\mu G}$]}]
			\pgfplotsset{cycle list shift=4}
			\addplot table [x={b0_fd_x}, y={b0_fd_y}] {data.dat};\label{plots:scan_norandom_5}
   	        \draw (6, \pgfkeysvalueof{/pgfplots/ymin}) -- (6, \pgfkeysvalueof{/pgfplots/ymax});

		\nextgroupplot[xlabel={$B_0$ [$\mathrm{\mu G}$]}, ymin=-200]
			\pgfplotsset{cycle list shift=5}
			\addplot table [x={b0_sync_Q_1.41_x}, y={b0_sync_Q_1.41_y}] {data.dat};\label{plots:scan_norandom_6}
    		\addplot table [x={b0_sync_U_1.41_x}, y={b0_sync_U_1.41_y}] {data.dat};\label{plots:scan_norandom_7}
    		\draw (6, \pgfkeysvalueof{/pgfplots/ymin}) -- (6, \pgfkeysvalueof{/pgfplots/ymax});

		\nextgroupplot[xlabel={$\psi_0$ [$^\circ$]}, ylabel={Log-Likelihood}]
			\addplot table [x={psi0_sync_I_0.408_x}, y={psi0_sync_I_0.408_y}] {data.dat};
			\addplot table [x={psi0_sync_Q_30_x}, y={psi0_sync_Q_30_y}] {data.dat};
			\addplot table [x={psi0_sync_U_30_x}, y={psi0_sync_U_30_y}] {data.dat};
			\addplot table [x={psi0_total_x}, y={psi0_total_y}] {data.dat};
			\draw (27, \pgfkeysvalueof{/pgfplots/ymin}) -- (27, \pgfkeysvalueof{/pgfplots/ymax});
			
		\nextgroupplot[xlabel={$\psi_0$ [$^\circ$]}]
			\pgfplotsset{cycle list shift=4}
			\addplot table [x={psi0_fd_x}, y={psi0_fd_y}] {data.dat};
			\draw (27, \pgfkeysvalueof{/pgfplots/ymin}) -- (27, \pgfkeysvalueof{/pgfplots/ymax});
			
		\nextgroupplot[xlabel={$\psi_0$ [$^\circ$]}]
			\pgfplotsset{cycle list shift=5}
			\addplot table [x={psi0_sync_Q_1.41_x}, y={psi0_sync_Q_1.41_y}] {data.dat};
    		\addplot table [x={psi0_sync_U_1.41_x}, y={psi0_sync_U_1.41_y}] {data.dat};
    		\draw (27, \pgfkeysvalueof{/pgfplots/ymin}) -- (27, \pgfkeysvalueof{/pgfplots/ymax});

		\nextgroupplot[xlabel={$\psi_1$ [$^\circ$]}, ylabel={Log-Likelihood}]
			\addplot table [x={psi1_sync_I_0.408_x}, y={psi1_sync_I_0.408_y}] {data.dat};
			\addplot table [x={psi1_sync_Q_30_x}, y={psi1_sync_Q_30_y}] {data.dat};
			\addplot table [x={psi1_sync_U_30_x}, y={psi1_sync_U_30_y}] {data.dat};
			\addplot table [x={psi1_total_x}, y={psi1_total_y}] {data.dat};
			\draw (7.95, \pgfkeysvalueof{/pgfplots/ymin}) -- (7.95, \pgfkeysvalueof{/pgfplots/ymax});

		\nextgroupplot[xlabel={$\psi_1$ [$^\circ$]}, scaled y ticks=base 10:-3]
			\pgfplotsset{cycle list shift=4}
			\addplot table [x={psi1_fd_x}, y={psi1_fd_y}] {data.dat};
			\draw (7.95, \pgfkeysvalueof{/pgfplots/ymin}) -- (7.95, \pgfkeysvalueof{/pgfplots/ymax});
			
		\nextgroupplot[xlabel={$\psi_1$ [$^\circ$]}]
			\pgfplotsset{cycle list shift=5}
			\addplot table [x={psi1_sync_Q_1.41_x}, y={psi1_sync_Q_1.41_y}] {data.dat};
    		\addplot table [x={psi1_sync_U_1.41_x}, y={psi1_sync_U_1.41_y}] {data.dat};
			\draw (7.95, \pgfkeysvalueof{/pgfplots/ymin}) -- (7.95, \pgfkeysvalueof{/pgfplots/ymax});

		\nextgroupplot[xlabel={$\chi_0$ [$^\circ$]}, ylabel={Log-Likelihood}, ymin=-4e6]
			\addplot table [x={chi0_sync_I_0.408_x}, y={chi0_sync_I_0.408_y}] {data.dat};
			\addplot table [x={chi0_sync_Q_30_x}, y={chi0_sync_Q_30_y}] {data.dat};
			\addplot table [x={chi0_sync_U_30_x}, y={chi0_sync_U_30_y}] {data.dat};
			\addplot table [x={chi0_total_x}, y={chi0_total_y}] {data.dat};
			\draw (25, \pgfkeysvalueof{/pgfplots/ymin}) -- (25, \pgfkeysvalueof{/pgfplots/ymax});

		\nextgroupplot[xlabel={$\chi_0$ [$^\circ$]}]
			\pgfplotsset{cycle list shift=4}
			\addplot table [x={chi0_fd_x}, y={chi0_fd_y}] {data.dat};
			\draw (25, \pgfkeysvalueof{/pgfplots/ymin}) -- (25, \pgfkeysvalueof{/pgfplots/ymax});

		\nextgroupplot[xlabel={$\chi_0$ [$^\circ$]}]
			\pgfplotsset{cycle list shift=5}
			\addplot table [x={chi0_sync_Q_1.41_x}, y={chi0_sync_Q_1.41_y}] {data.dat};
    		\addplot table [x={chi0_sync_U_1.41_x}, y={chi0_sync_U_1.41_y}] {data.dat};
			\draw (25, \pgfkeysvalueof{/pgfplots/ymin}) -- (25, \pgfkeysvalueof{/pgfplots/ymax});

		\coordinate (bot) at (rel axis cs:1,0);

	    \end{groupplot}

	            \path (top|-current bounding box.north)--
	                        coordinate(legendpos)
	                        (bot|-current bounding box.north);
	            \matrix[
	                    matrix of nodes,
	                    anchor=south,
	                    draw,
	                    inner sep=0.2em,
	                    draw
	                ]at([yshift=1ex]legendpos)
	                {
  	                    \ref{plots:scan_norandom_1}& Synchr. $408~\mathrm{MHz}$ ($I$)&[5pt]
                        \ref{plots:scan_norandom_2}& Synchr. $30~\mathrm{GHz}$ ($Q$)&[5pt]
  	                    \ref{plots:scan_norandom_3}& Synchr. $30~\mathrm{GHz}$ ($U$)&[5pt]
  	                    \ref{plots:scan_norandom_4}& Total&[5pt]\\
  	                    \ref{plots:scan_norandom_5}& Faraday depth&[5pt]
  	                    \ref{plots:scan_norandom_6}& Synchr. $1.41~\mathrm{GHz}$ ($Q$)&[5pt]
   	                    \ref{plots:scan_norandom_7}& Synchr. $1.41~\mathrm{GHz}$ ($U$)&[5pt]\\
  	                };
	    
	\end{tikzpicture}
\label{fig:parameter_scan_norandom}
\end{figure*}

After examining the one-dimensional parameter scans, we then check whether it is possible to infer the input parameters from the mock data set with simple minimization.
\ref{tab:mock_norandom_simplex} shows the values a Nelder-Mead minimizer \citep{Nelder:1965zz} yields when operating with the mock data set $\omega_1$. 
As mentioned above, only Faraday depth and synchrotron data at $408~\mathrm{MHz}$ and $30~\mathrm{GHz}$ were used according to the insights we drew from the parameter scans. 

\begin{table}
\caption{Log-likelihood maximizing parameter values for mock data $\omega_1$ inferred with a Nelder-Mead optimizer; showing the first significant digit of deviation.}
\begin{tabular}{@{} l | c c c c  @{}}
\toprule
    & $B_0~[\mathrm{\mu G}]$ & $\psi_0~[^\circ]$ & $\psi_1~[^\circ]$  & $\chi_0~[^\circ]$ \\ 
\midrule
 Mock values & $6.0$ & $27.0$ & $7.95$ & $25.0$ \\
 Reconstruction & $5.999$ & $26.99$ & $7.943$ & $25.003$\\
 \bottomrule
 \end{tabular}
 \label{tab:mock_norandom_simplex}
\end{table}

The minimizer is able to reliably find the correct parameter values, which suggests that the likelihood surface is well-behaved throughout the parameter space volume and not only along the optimum-intersecting axes. 
Note that in general it is advisable to use a gradient-free minimization scheme like Nelder-Mead due to possible non-smooth transitions that are particularly part of more complex magnetic field models. 

\begin{figure*}
\caption{\textbf{Mock data $\omega_1$:} Marginalized posterior plots and projected pairwise correlation plots from applying \pymultinest\ to mock data $\omega_1$. 
The dashed lines represent the $16\%$, $50\%$ and $84\%$ quantiles, respectively. 
The parameters for the mock data, indicated by the solid lines, were set to $B_0=6.0 ~\mathrm{\mu G}$, $\psi_0=27.0^\circ$, $\psi_1=7.95^\circ$, and $\chi_0=25^\circ$.
}
\centerline{\includegraphics[width=1.15\linewidth]{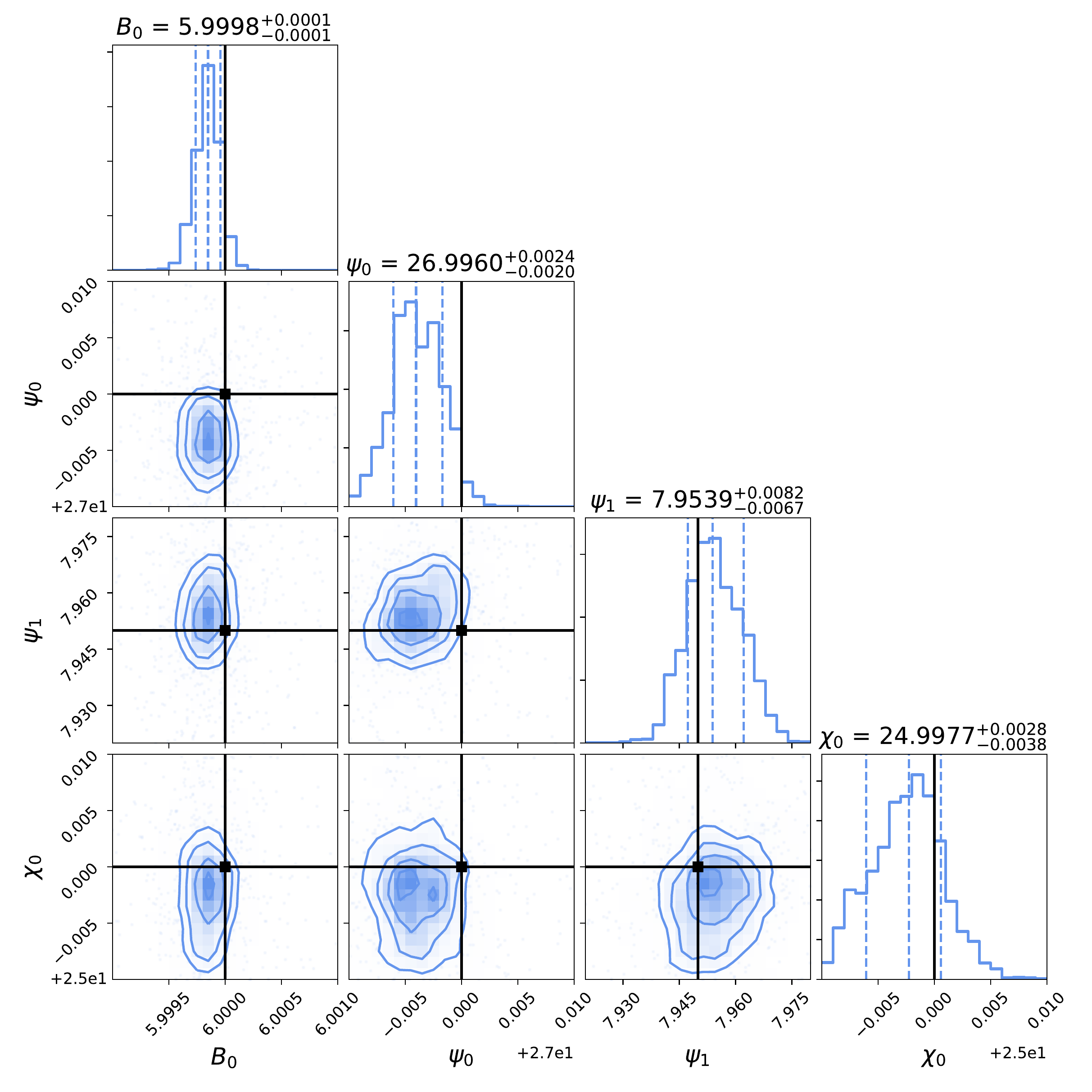}}
\label{fig:marginals_mock_norandom}
\end{figure*}

Finally, we use \pymultinest\ \citep{2014A&A...564A.125B} to explore the likelihood surface of the mock data $\omega_1$.
\ref{fig:marginals_mock_norandom} shows the marginalized probability density functions as well as pairwise correlation plots. 
As expected, $B_0$ is inferred with the highest precision; followed by $\psi_0$ and $\chi_0$, and finally $\psi_1$.  
In the course of this, the addition of mock noise causes the inferred parameter means to be shifted with respect to the true values. 
Different random seeds for the noise yield varying offsets. 
The likelihood is insensitive to these deviations, however, since we knew the true noise covariance matrix and take it into account. 
With the high signal-to-noise ratios, the $2\sigma$ intervals are narrow and cover the $\omega_1$'s \emph{true} parameter values. 
This means that the likelihood is consistent with the process of mock data creation and mock noise generation. 

\subsubsection{Regular and Random Magnetic Field}
In the following we repeat the steps from the previous section for mock data set $\omega_2$: scanning the parameter space, finding optimal parameter values with Nelder-Mead minimization and doing a full sampling with \pymultinest. 
\ref{fig:parameter_scan_random} shows that including a random magnetic field reduces the sensitivity of the ensemble likelihood considerably. 
In contrast to \ref{fig:parameter_scan_norandom}, now the log-likelihood values vary over one to three instead over eight orders of magnitude. 
As before, the signal for $B_0$ is strongest, followed by $\psi_0$ and $\chi_0$, and finally $\psi_1$. 
With respect to the parameters of the random magnetic field component we see that $\tau$, the parameter for the random magnetic field's strength, and $\alpha$, the random field's spectral index, exhibit a slight peak at their true values.
However, as foreseen in \ref{sec:likelihood}, $\tau$'s likelihood flattens significantly for large values. 
The fact that for $\alpha$ the likelihood has its maximum near the true mock data value is the incidental result of combining contrarily biased Faraday rotation and synchrotron radiation likelihoods. 
It should be noted that such shifts are not unexpected, since the mock data include a single realization of the Galactic and noise variance that can cause such chance alignment with slightly shifted parameters.
Finally, we note that Faraday rotation data would not be able to constrain $\tau$ and $\alpha$ reasonably. 
The total likelihood's shape around the true mock data value is rather flat for $\tau$ and $\alpha$. 
Their influence on the likelihood is comparably small in this mock scenario, but would increase with the strength of the random field component; here the setting is $B_0 = 6~\mathrm{\mu G}$ vs. $\tau = 2~\mathrm{\mu G}$. 
$\psi_1$, $\tau$, and $\alpha$ get traced by the observables -- at least slightly -- which is why we keep them for the further inference.
At this point the importance of this sensitivity analysis becomes evident, as we can draw the following conclusions:
If we find a parameter which has completely negligible or even misleading influence on the likelihood it should be excluded from inference. 
It would solely increase the dimensionality of the problem and with respect to minimizers and samplers behave in the best case as a noisy contribution and therefore disturb convergence. 

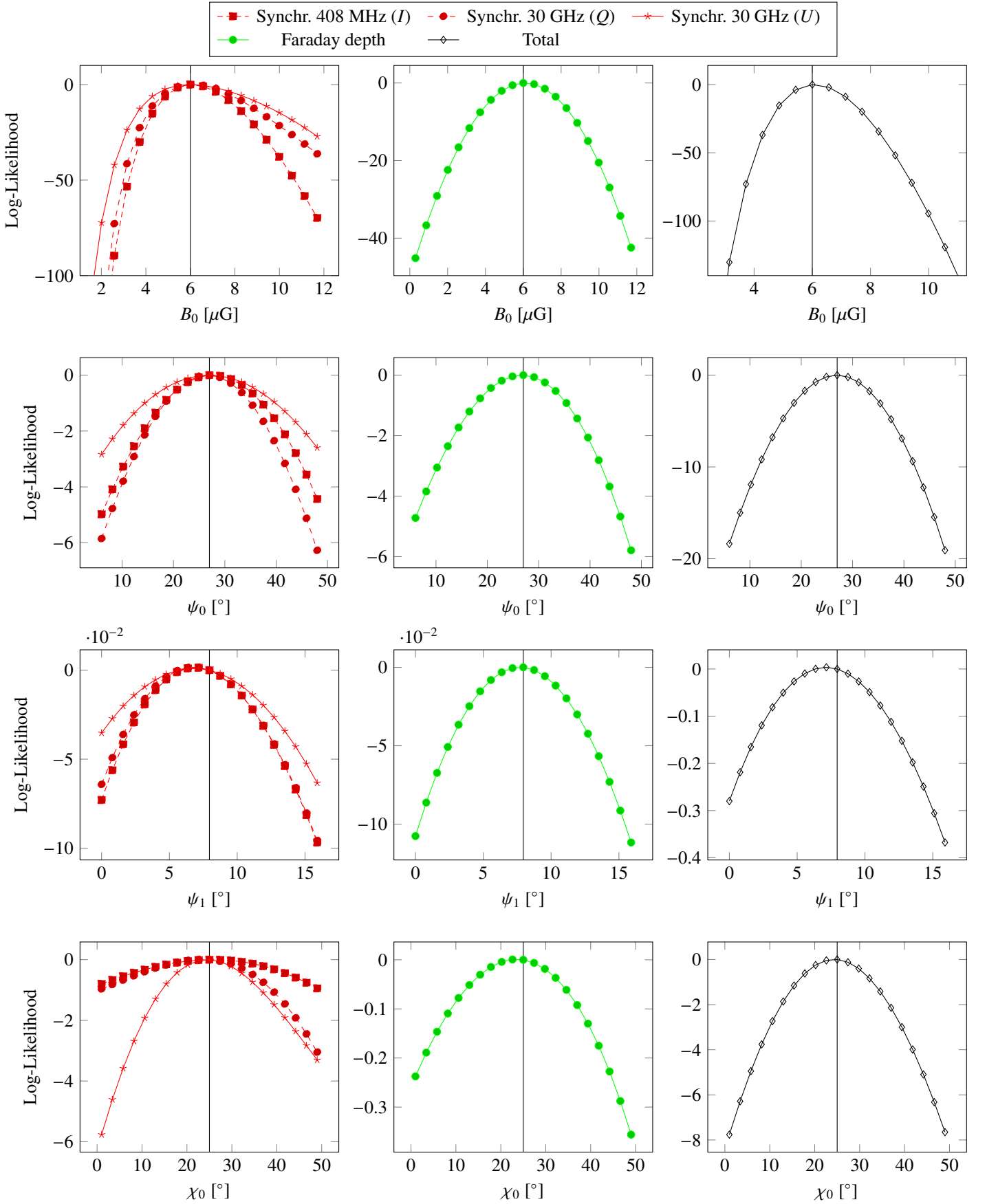
\begin{figure*}
\caption{\textbf{Mock data $\omega_2$:} Scans through the parameter space of the WMAP LSA regular plus isotropic random magnetic field model, including simulated additive measurement noise according to \citep{2016A&A...594A...1P}, \citet{2012A&A...542A..93O}. 
The mock data input parameter values are indicated by the vertical line.
Note that the plot of the \emph{total} likelihood does not include the synchrotron data at $1.41~\mathrm{GHz}$.}

	\begin{tikzpicture}
	\begin{groupplot}[group style = {group size = 3 by 4, horizontal sep = 30pt, vertical sep = 45pt}, width = 0.355\linewidth, table/search path={data/parameter_scan_random/}, cycle list name=longcolorlist]
		\nextgroupplot[xlabel={$B_0$ [$\mathrm{\mu G}$]}, ylabel={Log-Likelihood}, ymin=-100]
			\addplot table [x={b0_sync_I_0.408_x}, y={b0_sync_I_0.408_y}] {data.dat};\label{plots:scan_random_1}
			\addplot table [x={b0_sync_Q_30_x}, y={b0_sync_Q_30_y}] {data.dat};\label{plots:scan_random_2}
			\addplot table [x={b0_sync_U_30_x}, y={b0_sync_U_30_y}] {data.dat};\label{plots:scan_random_3}
   	        \draw (6, \pgfkeysvalueof{/pgfplots/ymin}) -- (6, \pgfkeysvalueof{/pgfplots/ymax});
		\coordinate (top) at (rel axis cs:0,1);

		\nextgroupplot[xlabel={$B_0$ [$\mathrm{\mu G}$]}]
			\pgfplotsset{cycle list shift=4}
			\addplot table [x={b0_fd_x}, y={b0_fd_y}] {data.dat};\label{plots:scan_random_5}
   	        \draw (6, \pgfkeysvalueof{/pgfplots/ymin}) -- (6, \pgfkeysvalueof{/pgfplots/ymax});

		\nextgroupplot[xlabel={$B_0$ [$\mathrm{\mu G}$]}, ymin=-140]
			\pgfplotsset{cycle list shift=3}
			\addplot table [x={b0_total_x}, y={b0_total_y}] {data.dat};\label{plots:scan_random_4}
    		\draw (6, \pgfkeysvalueof{/pgfplots/ymin}) -- (6, \pgfkeysvalueof{/pgfplots/ymax});

		\nextgroupplot[xlabel={$\psi_0$ [$^\circ$]}, ylabel={Log-Likelihood}]
			\addplot table [x={psi0_sync_I_0.408_x}, y={psi0_sync_I_0.408_y}] {data.dat};
			\addplot table [x={psi0_sync_Q_30_x}, y={psi0_sync_Q_30_y}] {data.dat};
			\addplot table [x={psi0_sync_U_30_x}, y={psi0_sync_U_30_y}] {data.dat};
			\draw (27, \pgfkeysvalueof{/pgfplots/ymin}) -- (27, \pgfkeysvalueof{/pgfplots/ymax});
			
		\nextgroupplot[xlabel={$\psi_0$ [$^\circ$]}]
			\pgfplotsset{cycle list shift=4}
			\addplot table [x={psi0_fd_x}, y={psi0_fd_y}] {data.dat};
			\draw (27, \pgfkeysvalueof{/pgfplots/ymin}) -- (27, \pgfkeysvalueof{/pgfplots/ymax});
			
		\nextgroupplot[xlabel={$\psi_0$ [$^\circ$]}]
			\pgfplotsset{cycle list shift=3}
			\addplot table [x={psi0_total_x}, y={psi0_total_y}] {data.dat};
    		\draw (27, \pgfkeysvalueof{/pgfplots/ymin}) -- (27, \pgfkeysvalueof{/pgfplots/ymax});

		\nextgroupplot[xlabel={$\psi_1$ [$^\circ$]}, ylabel={Log-Likelihood}, scaled y ticks=base 10:2]
			\addplot table [x={psi1_sync_I_0.408_x}, y={psi1_sync_I_0.408_y}] {data.dat};
			\addplot table [x={psi1_sync_Q_30_x}, y={psi1_sync_Q_30_y}] {data.dat};
			\addplot table [x={psi1_sync_U_30_x}, y={psi1_sync_U_30_y}] {data.dat};
			\draw (7.95, \pgfkeysvalueof{/pgfplots/ymin}) -- (7.95, \pgfkeysvalueof{/pgfplots/ymax});

		\nextgroupplot[xlabel={$\psi_1$ [$^\circ$]}, scaled y ticks=base 10:2]
			\pgfplotsset{cycle list shift=4}
			\addplot table [x={psi1_fd_x}, y={psi1_fd_y}] {data.dat};
			\draw (7.95, \pgfkeysvalueof{/pgfplots/ymin}) -- (7.95, \pgfkeysvalueof{/pgfplots/ymax});
			
		\nextgroupplot[xlabel={$\psi_1$ [$^\circ$]}]
			\pgfplotsset{cycle list shift=3}
			\addplot table [x={psi1_total_x}, y={psi1_total_y}] {data.dat};			
			\draw (7.95, \pgfkeysvalueof{/pgfplots/ymin}) -- (7.95, \pgfkeysvalueof{/pgfplots/ymax});

		\nextgroupplot[xlabel={$\chi_0$ [$^\circ$]}, ylabel={Log-Likelihood}]
			\addplot table [x={chi0_sync_I_0.408_x}, y={chi0_sync_I_0.408_y}] {data.dat};
			\addplot table [x={chi0_sync_Q_30_x}, y={chi0_sync_Q_30_y}] {data.dat};
			\addplot table [x={chi0_sync_U_30_x}, y={chi0_sync_U_30_y}] {data.dat};
			\draw (25, \pgfkeysvalueof{/pgfplots/ymin}) -- (25, \pgfkeysvalueof{/pgfplots/ymax});

		\nextgroupplot[xlabel={$\chi_0$ [$^\circ$]}]
			\pgfplotsset{cycle list shift=4}
			\addplot table [x={chi0_fd_x}, y={chi0_fd_y}] {data.dat};
			\draw (25, \pgfkeysvalueof{/pgfplots/ymin}) -- (25, \pgfkeysvalueof{/pgfplots/ymax});

		\nextgroupplot[xlabel={$\chi_0$ [$^\circ$]}]
			\pgfplotsset{cycle list shift=3}
			\addplot table [x={chi0_total_x}, y={chi0_total_y}] {data.dat};
			\draw (25, \pgfkeysvalueof{/pgfplots/ymin}) -- (25, \pgfkeysvalueof{/pgfplots/ymax});

		\coordinate (bot) at (rel axis cs:1,0);

	    \end{groupplot}

	            \path (top|-current bounding box.north)--
	                        coordinate(legendpos)
	                        (bot|-current bounding box.north);
	            \matrix[
	                    matrix of nodes,
	                    anchor=south,
	                    draw,
	                    inner sep=0.2em,
	                    draw
	                ]at([yshift=1ex]legendpos)
	                {
  	                    \ref{plots:scan_random_1}& Synchr. $408~\mathrm{MHz}$ ($I$)&[5pt]
                        \ref{plots:scan_random_2}& Synchr. $30~\mathrm{GHz}$ ($Q$)&[5pt]
  	                    \ref{plots:scan_random_3}& Synchr. $30~\mathrm{GHz}$ ($U$)&[5pt]\\
  	                    \ref{plots:scan_random_5}& Faraday depth&[5pt]
   	                    \ref{plots:scan_random_4}& Total&[5pt]\\
  	                };
	    
	\end{tikzpicture}
\label{fig:parameter_scan_random}
\end{figure*}

\begin{figure*}\ContinuedFloat

	\begin{tikzpicture}
	\begin{groupplot}[group style = {group size = 3 by 4, horizontal sep = 30pt, vertical sep = 45pt}, width = 0.355\linewidth, table/search path={data/parameter_scan_random/}, cycle list name=longcolorlist]
			
		\nextgroupplot[xlabel={$\tau$ [$\mathrm{\mu G}$]}, ylabel={Log-Likelihood}, ymin=-2]
			\addplot table [x={random_rms_sync_I_0.408_x}, y={random_rms_sync_I_0.408_y}] {data.dat};
			\addplot table [x={random_rms_sync_Q_30_x}, y={random_rms_sync_Q_30_y}] {data.dat};
			\addplot table [x={random_rms_sync_U_30_x}, y={random_rms_sync_U_30_y}] {data.dat};
			\draw (2, \pgfkeysvalueof{/pgfplots/ymin}) -- (2, \pgfkeysvalueof{/pgfplots/ymax});
			
		\nextgroupplot[xlabel={$\tau$ [$\mathrm{\mu G}$]}, ymin=-2]
			\pgfplotsset{cycle list shift=4}
			\addplot table [x={random_rms_fd_x}, y={random_rms_fd_y}] {data.dat};
			\draw (2, \pgfkeysvalueof{/pgfplots/ymin}) -- (2, \pgfkeysvalueof{/pgfplots/ymax});
			
		\nextgroupplot[xlabel={$\tau$ [$\mathrm{\mu G}$]}, ymin=-4]
			\pgfplotsset{cycle list shift=3}
			\addplot table [x={random_rms_total_x}, y={random_rms_total_y}] {data.dat};
    		\draw (2, \pgfkeysvalueof{/pgfplots/ymin}) -- (2, \pgfkeysvalueof{/pgfplots/ymax});

		\nextgroupplot[xlabel={$\alpha$}, ylabel={Log-Likelihood}]
			\addplot table [x={random_a0_sync_I_0.408_x}, y={random_a0_sync_I_0.408_y}] {data.dat};
			\addplot table [x={random_a0_sync_Q_30_x}, y={random_a0_sync_Q_30_y}] {data.dat};
			\addplot table [x={random_a0_sync_U_30_x}, y={random_a0_sync_U_30_y}] {data.dat};
			\draw (1.7, \pgfkeysvalueof{/pgfplots/ymin}) -- (1.7, \pgfkeysvalueof{/pgfplots/ymax});

		\nextgroupplot[xlabel={$\alpha$}]
			\pgfplotsset{cycle list shift=4}
			\addplot table [x={random_a0_fd_x}, y={random_a0_fd_y}] {data.dat};
			\draw (1.7, \pgfkeysvalueof{/pgfplots/ymin}) -- (1.7, \pgfkeysvalueof{/pgfplots/ymax});

		\nextgroupplot[xlabel={$\alpha$}]
			\pgfplotsset{cycle list shift=3}
			\addplot table [x={random_a0_total_x}, y={random_a0_total_y}] {data.dat};			
			\draw (1.7, \pgfkeysvalueof{/pgfplots/ymin}) -- (1.7, \pgfkeysvalueof{/pgfplots/ymax});

		\coordinate (bot) at (rel axis cs:1,0);

	    \end{groupplot}
	    
	            \path (top|-current bounding box.north)--
	                        coordinate(legendpos)
	                        (bot|-current bounding box.north);
	            \matrix[
	                    matrix of nodes,
	                    anchor=south,
	                    draw,
	                    inner sep=0.2em,
	                    draw
	                ]at([yshift=1ex]legendpos)
	                {
  	                    \ref{plots:scan_random_1}& Synchr. $408~\mathrm{MHz}$ ($I$)&[5pt]
                        \ref{plots:scan_random_2}& Synchr. $30~\mathrm{GHz}$ ($Q$)&[5pt]
  	                    \ref{plots:scan_random_3}& Synchr. $30~\mathrm{GHz}$ ($U$)&[5pt]\\
  	                    \ref{plots:scan_random_5}& Faraday depth&[5pt]
  	                    \ref{plots:scan_random_4}& Total&[5pt]\\
  	                };
	    
	\end{tikzpicture}
\end{figure*}

For completeness, we visualize the importance of the regularizing determinant in \ref{eq:ensemble_likelihood}. 
\ref{fig:parameter_scan_random_nodet} shows that without the determinant the ensemble likelihood favors too high random field strengths and spectral indices. 

\begin{figure*}
\caption{\textbf{Mock data $\omega_2$, without determinant term:} Scans through the parameter space of the WMAP LSA regular plus isotropic random magnetic field model, including simulated additive measurement noise according to \citet{2006A&A...448..411W}, \citep{2016A&A...594A...1P}, \citet{2012A&A...542A..93O}. 
For these plots the ensemble likelihood was evaluated without the determinant term.
The mock data input parameter values are indicated by the vertical line. 
Note that the plot of the \emph{total} likelihood does not include the synchrotron data at $1.41~\mathrm{GHz}$.}

	\begin{tikzpicture}
	\begin{groupplot}[group style = {group size = 3 by 4, horizontal sep = 30pt, vertical sep = 45pt}, width = 0.355\linewidth, table/search path={data/parameter_scan_random_nodet/}, cycle list name=longcolorlist]
		\nextgroupplot[xlabel={$\tau$ [$\mathrm{\mu G}$]}, ylabel={Log-Likelihood}, ymin=-2]
			\addplot table [x={random_rms_sync_I_0.408_x}, y={random_rms_sync_I_0.408_y}] {data.dat};\label{plots:scan_random_nodet_1}
			\addplot table [x={random_rms_sync_Q_30_x}, y={random_rms_sync_Q_30_y}] {data.dat};\label{plots:scan_random_nodet_2}
			\addplot table [x={random_rms_sync_U_30_x}, y={random_rms_sync_U_30_y}] {data.dat};\label{plots:scan_random_nodet_3}
			\draw (2, \pgfkeysvalueof{/pgfplots/ymin}) -- (2, \pgfkeysvalueof{/pgfplots/ymax});
			
		\nextgroupplot[xlabel={$\tau$ [$\mathrm{\mu G}$]}, ymin=-2]
			\pgfplotsset{cycle list shift=4}
			\addplot table [x={random_rms_fd_x}, y={random_rms_fd_y}] {data.dat};
			\draw (2, \pgfkeysvalueof{/pgfplots/ymin}) -- (2, \pgfkeysvalueof{/pgfplots/ymax});\label{plots:scan_random_nodet_5}
			
		\nextgroupplot[xlabel={$\tau$ [$\mathrm{\mu G}$]}, ymin=-4]
			\pgfplotsset{cycle list shift=3}
			\addplot table [x={random_rms_total_x}, y={random_rms_total_y}] {data.dat};\label{plots:scan_random_nodet_4}
    		\draw (2, \pgfkeysvalueof{/pgfplots/ymin}) -- (2, \pgfkeysvalueof{/pgfplots/ymax});

		\nextgroupplot[xlabel={$\alpha$}, ylabel={Log-Likelihood}]
			\addplot table [x={random_a0_sync_I_0.408_x}, y={random_a0_sync_I_0.408_y}] {data.dat};
			\addplot table [x={random_a0_sync_Q_30_x}, y={random_a0_sync_Q_30_y}] {data.dat};
			\addplot table [x={random_a0_sync_U_30_x}, y={random_a0_sync_U_30_y}] {data.dat};
			\draw (1.7, \pgfkeysvalueof{/pgfplots/ymin}) -- (1.7, \pgfkeysvalueof{/pgfplots/ymax});

		\nextgroupplot[xlabel={$\alpha$}]
			\pgfplotsset{cycle list shift=4}
			\addplot table [x={random_a0_fd_x}, y={random_a0_fd_y}] {data.dat};
			\draw (1.7, \pgfkeysvalueof{/pgfplots/ymin}) -- (1.7, \pgfkeysvalueof{/pgfplots/ymax});

		\nextgroupplot[xlabel={$\alpha$}]
			\pgfplotsset{cycle list shift=3}
			\addplot table [x={random_a0_total_x}, y={random_a0_total_y}] {data.dat};			
			\draw (1.7, \pgfkeysvalueof{/pgfplots/ymin}) -- (1.7, \pgfkeysvalueof{/pgfplots/ymax});

		\coordinate (bot) at (rel axis cs:1,0);

	    \end{groupplot}
	    
	            \path (top|-current bounding box.north)--
	                        coordinate(legendpos)
	                        (bot|-current bounding box.north);
	            \matrix[
	                    matrix of nodes,
	                    anchor=south,
	                    draw,
	                    inner sep=0.2em,
	                    draw
	                ]at([yshift=1ex]legendpos)
	                {
  	                    \ref{plots:scan_random_nodet_1}& Synchr. $408~\mathrm{MHz}$ ($I$)&[5pt]
                        \ref{plots:scan_random_nodet_2}& Synchr. $30~\mathrm{GHz}$ ($Q$)&[5pt]
  	                    \ref{plots:scan_random_nodet_3}& Synchr. $30~\mathrm{GHz}$ ($U$)&[5pt]\\
  	                    \ref{plots:scan_random_nodet_5}& Faraday depth&[5pt]
  	                    \ref{plots:scan_random_nodet_4}& Total&[5pt]\\
  	                };
	    
	\end{tikzpicture}
	\label{fig:parameter_scan_random_nodet}
	\end{figure*}

As in \ref{sec:regular_mock_tests}, we continue by inferring the parameter values of $\omega_2$ using a Nelder-Mead minimizer. 
\ref{tab:mock_random_simplex} shows the results of the optimization whereby we see that as expected the accuracy is significantly lower than without a random magnetic field component, cf.\ \ref{tab:mock_norandom_simplex}. 
One can also see the trend that $\tau$ gets overestimated, which already became apparent in the parameter scan, cf.\ \ref{fig:parameter_scan_random}. 

\begin{table*}
\caption{Log-likelihood maximizing parameter values for mock data $\omega_2$ inferred with a Nelder-Mead optimizer; showing two significant digits of deviation.}
\begin{center}
\begin{tabular}{@{} l | c c c c c c @{}}
\toprule
    & $B_0~[\mathrm{\mu G}]$ & $\psi_0~[^\circ]$ & $\psi_1~[^\circ]$  & $\chi_0~[^\circ]$ & $\tau~[\mathrm{\mu G}]$ & $\alpha$\\ 
\midrule
 Mock values & $6.0$ & $27.0$ & $7.95$ & $25.0$ & $2.0$ & $1.7$ \\
 Reconstruction & $6.063$ & $26.86$ & $8.38$ & $23.9$ & $2.11$ & $1.698$\\
 \bottomrule
 \end{tabular}
\end{center}
 \label{tab:mock_random_simplex}
\end{table*}

\begin{figure*}
\caption{\textbf{Mock data $\omega_2$:} Marginalized posterior plots and projected pairwise correlation plots from applying \pymultinest\ to mock data $\omega_2$. 
The dashed lines represent the $16\%$, $50\%$ and $84\%$ quantiles, respectively. 
The parameters for the mock data, indicated by the solid lines, were set to $B_0=6.0 ~\mathrm{\mu G}$, $\psi_0=27.0^\circ$, $\psi_1=7.95^\circ$, $\chi_0=25^\circ$, $\tau=2~\mathrm{\mu G}$, and $\alpha = 1.7$.
}
\centerline{\includegraphics[width=1.15\linewidth]{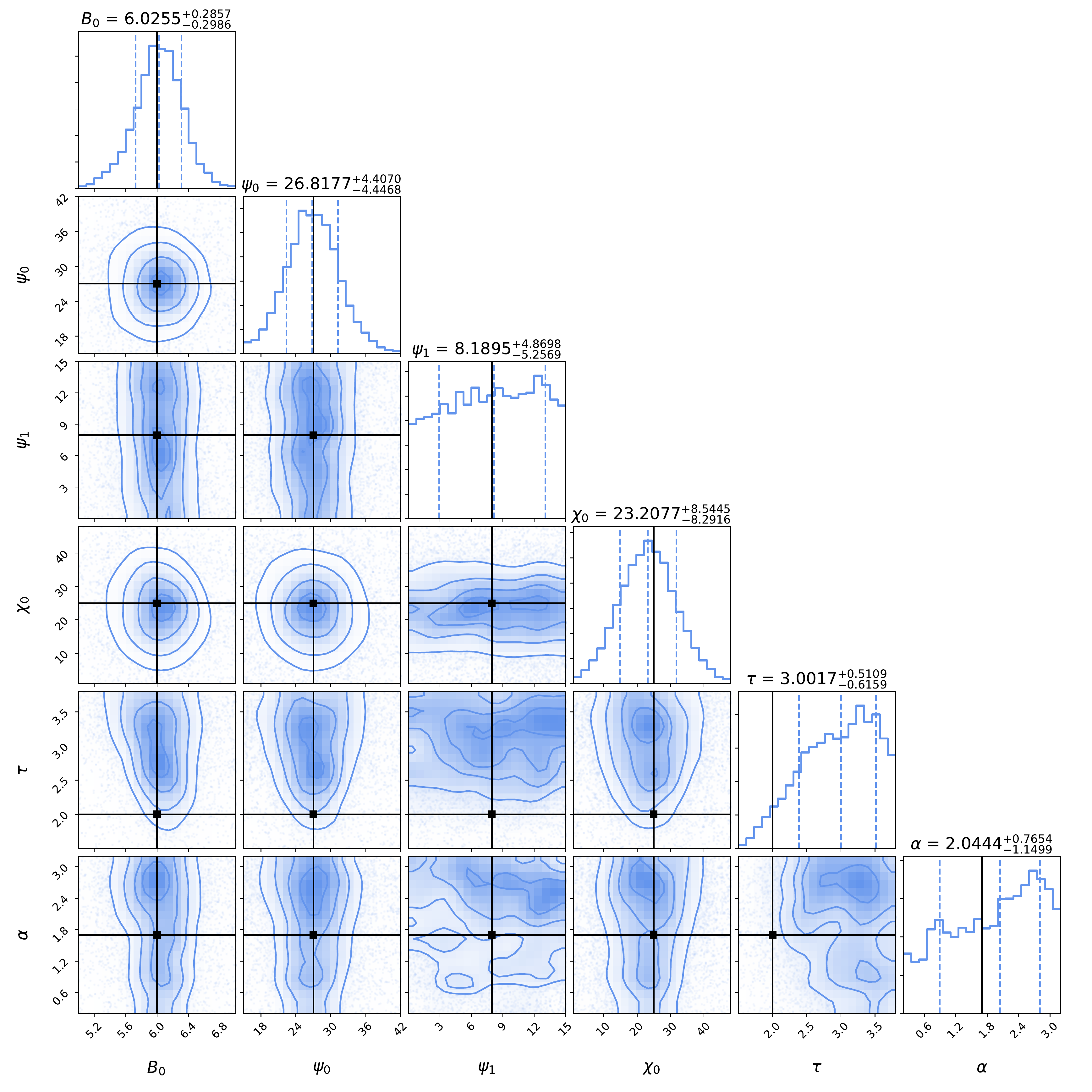}}
\label{fig:marginals_mock_random}
\end{figure*}

Finally, \ref{fig:marginals_mock_random} shows the marginal plots based on a \pymultinest\ run on the mock data set $\omega_2$. 
First, we recognize that the Galactic variance caused rather broad uncertainties. 
Nevertheless, the uncertainty intervals are highly reasonable: for example, although the sample mean value for $B_0$ lies rather precisely at $6~\mathrm{\mu G}$, the maximum likelihood value is significantly shifted to the right. 
Furthermore, one sees that the Galactic variance washes out almost all predictive power on $\psi_1$.
The fact that the sample mean matches the mock data's generating parameter is mainly due to the fact that the true value is at the center of the prior volume. 
As seen before, $\tau$ gets overestimated, while $2~\mathrm{\mu G}$ still lies within the $2\sigma$ interval. 
Interestingly enough, looking at the joint probability density plot for $\tau$ and $\alpha$, one sees that for larger $\tau$ also larger $\alpha$ become more likely. 
This is on the one hand an indicator for an unsurprising degeneracy between the total strength and the spectral index. 
On the other hand, we expect that the predictive power on one of the parameters can be increased by fixing the other by the use of strong prior information. 

\begin{figure*}
\caption{\textbf{Mock data $\omega_2$ in combination with a $\chi^2$ likelihood:} Marginalized posterior plots and projected pairwise correlation plots from applying \pymultinest\ to mock data $\omega_2$ using a simple $\chi^2$ likelihood that not reflects the influence of the Galactic variance. 
The dashed lines represent the $16\%$, $50\%$ and $84\%$ quantiles, respectively. 
The parameters for the mock data, indicated by the solid lines, were set to $B_0=6.0 ~\mathrm{\mu G}$, $\psi_0=27.0^\circ$, $\psi_1=7.95^\circ$, $\chi_0=25^\circ$, $\tau=2~\mathrm{\mu G}$, and $\alpha = 1.7$.
}
\centerline{\includegraphics[width=1.15\linewidth]{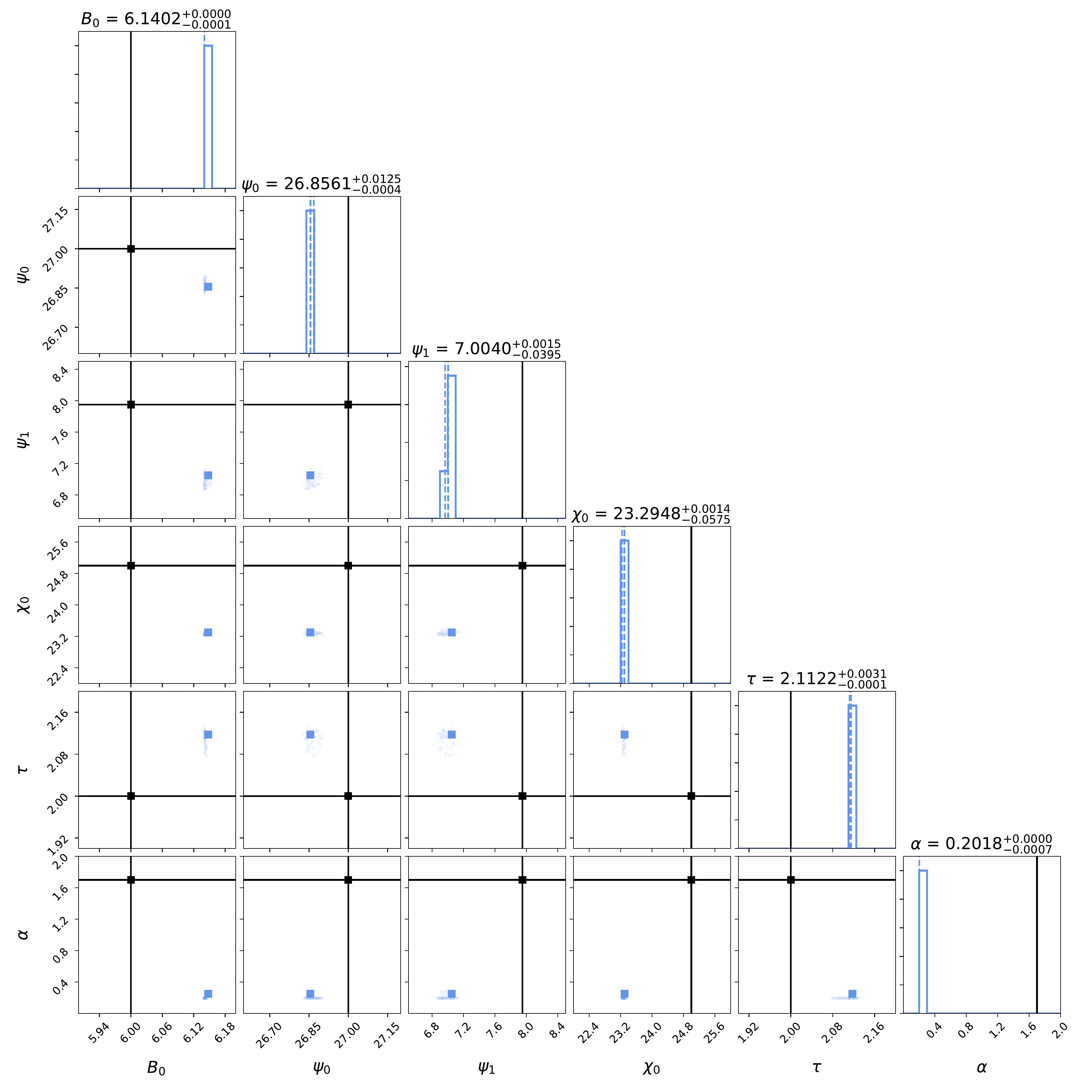}}
\label{fig:marginals_mock_random_simple}
\end{figure*}

Note, that a naive $\chi^2$ likelihood which, unlike the ensemble likelihood, does not reflect the Galactic variance massively underestimates the uncertainties introduced by the random magnetic field. 
\ref{fig:marginals_mock_random_simple} shows the result of \pymultinest\ maximizing a $\chi^2$ likelihood on the $\omega_2$ mock data set.
The spectral index is pushed to a small value of $\alpha = 0.202$ making the random magnetic field rather white. 
As a consequence, in the ensemble mean the influence of the random magnetic field maximally cancels out as the set of samples in the ensemble is finite. 
The other parameters then heavily over-fit the variations in the mock-data which come from its specific random magnetic field realization. 
This illustrates the importance of taking the Galactic variance into account when doing model parameter inference. 

\subsubsection{Model Comparison}
One strength of sampling methods like \multinest\ is that they produce an estimate for the \emph{evidence}. 
As discussed in \ref{sec:parameter_inference_model_comparison}, the evidence is crucial for model selection. 
To illustrate the procedure, we set up the following scenario:
Given the prevailing mock data set $\omega_2$, we compare two models that are both trivial versions of the WMAP LSA model. 
The only free parameter is now $B_0$.
For model $M_1$ the values for the \emph{hidden} parameters are equal to those of the mock data, while for model $M_2$ they are fixed to $\psi_0 = 3.0^\circ$, $\psi_1 = 25.0^\circ$, and $\chi_0 = 7.0^\circ$.
\ref{tab:mock_pymultinest} shows that the log-evidence for $M_1$ is significantly higher than for $M_2$, corresponding to a massive Bayes factor of $R=2.47\cdot 10^{10}$. 
But besides the quality of fit the evidence also takes the model's complexity into account. 
The fewer parameters a model has, the smaller is its total parameter space volume. 
Hence, even if a rather complicated model has a better best-fit estimate than a simpler one, if over-fitting occurs its evidence value will be worse.
\ref{tab:mock_pymultinest} also shows the log-evidence for the full four-parameter WMAP LSA model ($M_0$).
The log-evidence for $M_0$ lies in between those of $M_1$ and $M_2$; the Bayes factor between $M_0$ and $M_1$ is $R=4.18$, which means that there is \emph{substantial} evidence that $M_1$ is more likely \citep{jeffreys1998theory}. 
Thus, one sees the penalty coming from $M_0$'s larger parameter space volume compared to $M_1$. 
However, the improvements of a better parameter fit may compensate for this penalty as the comparison with $M_2$ shows. 

\begin{table*}
\caption{Sample mean and log-evidence values from \pymultinest\ for different WMAP LSA plus random magnetic field models based on the mock data set $\omega_2$.
At $M_0$ all four WMAP parameters are flexible; at $M_1$ and $M_2$ only $B_0$ is adjustable. 
An asterisk ($^*$) indicates a fixed value.
In any case, the random magnetic field's parameters were kept at their mock data's default value $\tau = 2~\mathrm{\mu G}$, and $\alpha=1.7$
}
\centering
\begin{tabular}{@{} l | c c c c c @{}}
\toprule
    & Log-Evidence & $B_0~[\mathrm{\mu G}]$ & $\psi_0~[^\circ]$ & $\psi_1~[^\circ]$  & $\chi_0~[^\circ]$ \\ 
\midrule
 Mock values 	& 					& $6.0$ 			& $27.0$ 			& $7.95$ 			& $25.0$ 		\\
 $M_0$ 			& $13.66 \pm 0.20$ 	& $6.10 \pm 0.253$ 	& $26.89 \pm 3.42$ 	&  $8.43 \pm 4.39$ 	& $23.92 \pm 6.12$ 	 \\
 $M_1$ 			& $15.09 \pm 0.18$ 	& $6.10 \pm 0.212$ 	& $27.0^*$ 			&  $7.95^*$ 		& $25.0^*$ 	 \\
 $M_2$ 			& $-8.84 \pm 0.15$ & $6.229 \pm 0.232$ 	& $3.0^*$ 			&  $25.0^*$ 		& $7.0^*$ 	 \\
 \bottomrule
 \end{tabular}
 \label{tab:mock_pymultinest}
\end{table*}

\subsection{Application to Real Data}
\begin{figure*}
\caption{\textbf{Synchrotron data, purely ordered WMAP LSA magnetic field} Marginalized posterior plots and projected pairwise correlation plots from applying \pymultinest\ to $408~\mathrm{MHz}$ and $30~\mathrm{GHz}$ synchrotron data from \citet{2016A&A...594A...1P}. 
The dashed lines represent the $16\%$, $50\%$ and $84\%$ quantiles, respectively. 
}
\centerline{\includegraphics[width=1.15\linewidth]{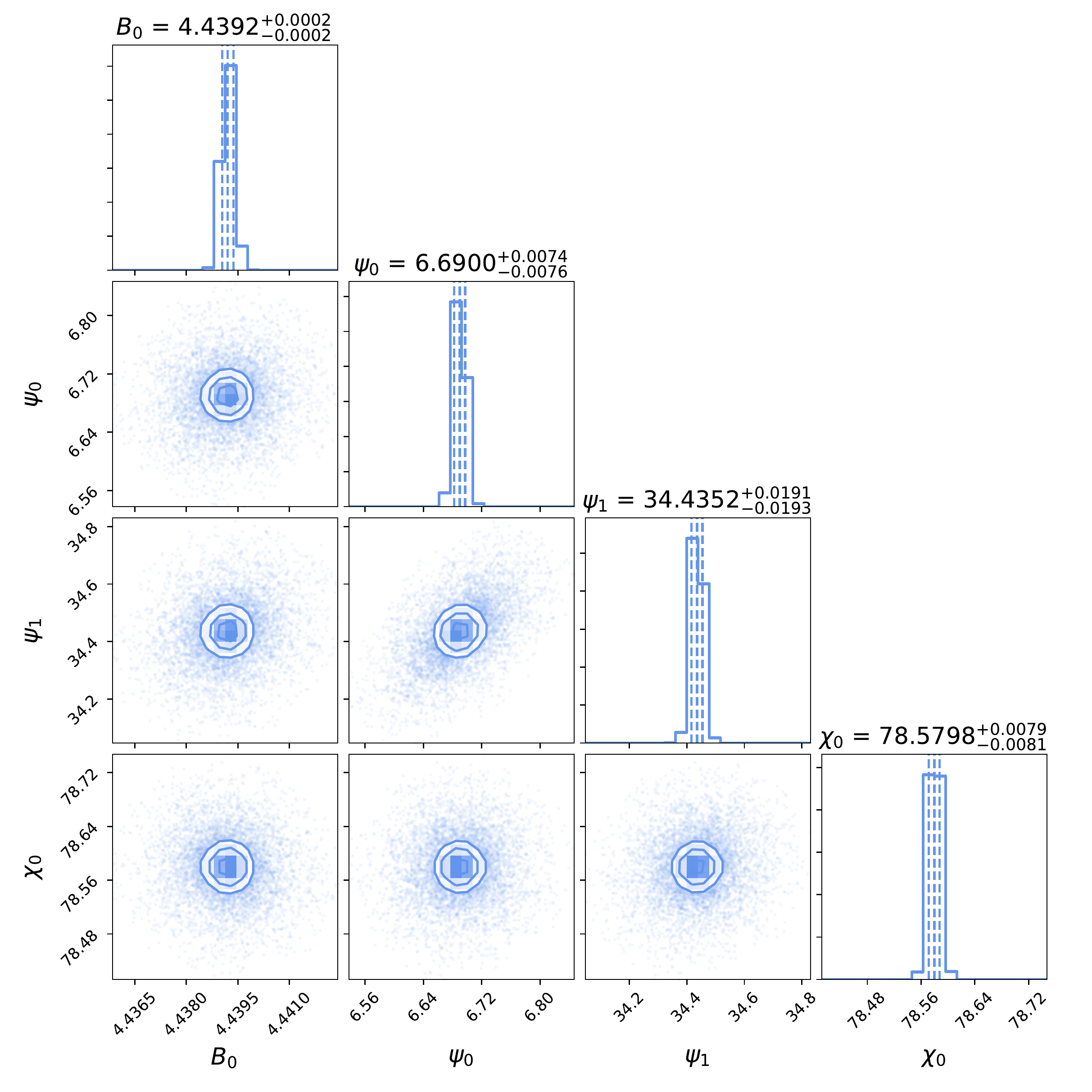}}
\label{fig:marginals_data_norandom_sync}
\end{figure*}

In \ref{sec:mock_data_tests}, we verified that the \imagine\ pipeline produces self-consistent results for the WMAP LSA model in combination with the chosen observables. 
Now we analyze the likelihood structure for the real synchrotron data at $408~\mathrm{MHz}$ and $30~\mathrm{GHz}$ \citep{2016A&A...594A...1P}, and the Faraday depth data \citep{2012A&A...542A..93O}\footnote{For using \imagine\ in production it is advisable to use the raw data compiled by \citet{2012A&A...542A..93O} as this ensures that there is no alteration of the noise information by a Wiener filter.}
Generally, it is advisable to thoroughly prepare the input data by masking regions in the sky obviously perturbed by local phenomena, for example supernova remnants. 
However, for this paper this is beyond the scope, as the goal is to illustrate the concepts behind \imagine\ rather than producing high-precision estimates. 
First, we use the synchrotron data to constrain the parameters of the purely ordered WMAP LSA model; so far no random fields are included in the magnetic field nor in the likelihood. 
The result is shown in \ref{fig:marginals_data_norandom_sync}. 

The resulting uncertainties are quantitatively consistent with the mock-data results, cf.\ \ref{fig:marginals_mock_norandom}.
Qualitatively speaking, $B_0$ is determined with the highest accuracy, followed by $\psi_0$ and $\chi_0$, and finally $\psi_1$. 
Also the inferred magnetic field strength $B_0 = 4.44~\mathrm{\mu G}$ is of a reasonable order of magnitude \citep{2010A&A...522A..73R, 2006ChJAS...6b.211H}. 
$\psi_0 = 6.69^\circ$ lies within the wide range of estimates one can find in literature, e.g., $8^\circ$ \citep{Beck2001, 2006ChJAS...6b.211H} and $35^\circ$ \citep{page:2007}. 
However, $\psi_1=34.4^\circ$ and $\chi_0=78.6^\circ$ are far off from the best-fit values given in \citet{page:2007}, namely $\psi_1=0.9^\circ$ and $\chi_0=25^\circ$. 
This, in combination with the very small uncertainties, indicates that an inference neglecting the influence of random components in the magnetic field as well as the likelihood is making matters too easy. 

\begin{figure*}
\caption{\textbf{Faraday rotation data, purely ordered WMAP LSA magnetic field} Marginalized posterior plots and projected pairwise correlation plots from applying \pymultinest\ to Faraday rotation data from \citet{2012A&A...542A..93O}. 
The dashed lines represent the $16\%$, $50\%$ and $84\%$ quantiles, respectively. 
}
\centerline{\includegraphics[width=1.15\linewidth]{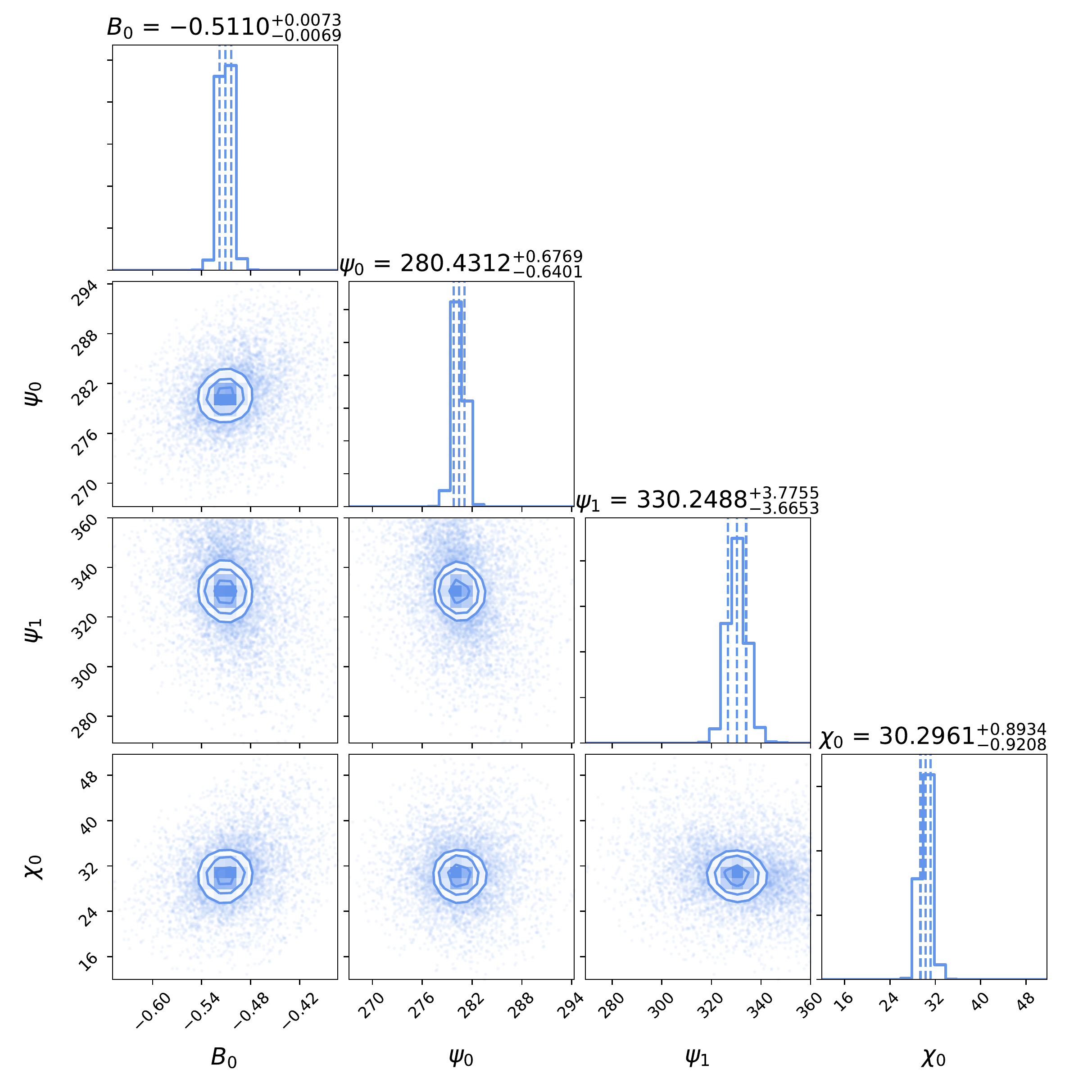}}
\label{fig:marginals_data_norandom_fd}
\end{figure*}

\begin{figure*}
\caption{Comparison of Faraday depth maps in $\mathrm{rad/m^2}$.}
\begin{subfigure}{1\textwidth}
  \centering
  	\includegraphics[width=.5\linewidth]{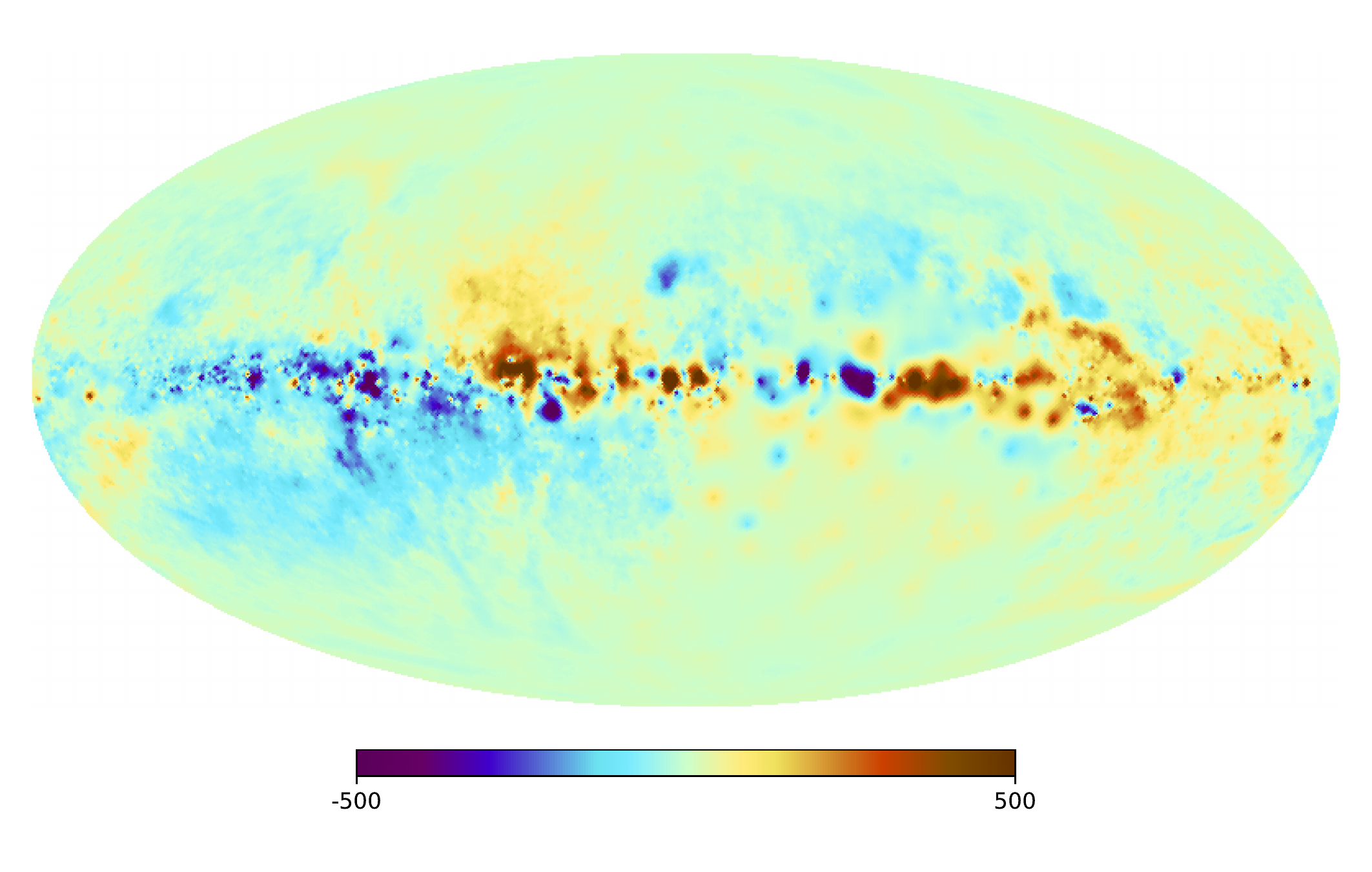}
  \caption{Map of the Galactic Faraday depth given by \citet{2012A&A...542A..93O} in $\mathrm{rad/m^2}$.}
\end{subfigure}\\
\centering
\begin{subfigure}{0.495\textwidth}
  \centering
  \captionsetup{width=.9\linewidth}
	  \includegraphics[width=\linewidth]{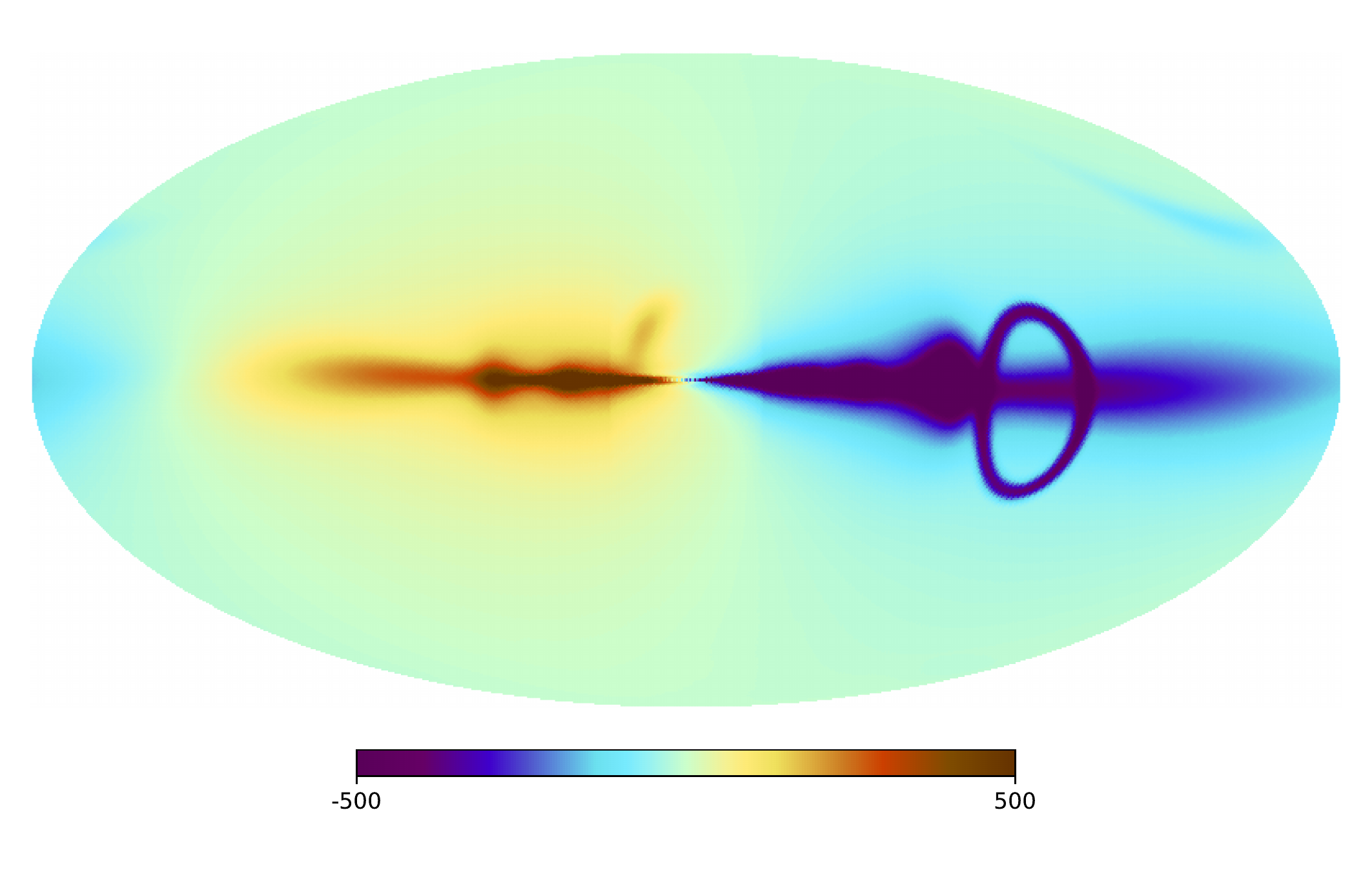}
  \caption{Map of the Galactic Faraday depth produced with \hammurabiX\ based on the WMAP LSA model in $\mathrm{rad/m^2}$. The model's parameters were set to $B_0=1.5~\mathrm{\mu G}$, $\psi_0=27.0^\circ$, $\psi_1=0.9^\circ$, and $\chi_0=25.0^\circ$, following \citet{page:2007}.}
\end{subfigure}
\begin{subfigure}{0.495\textwidth}
  \centering
   \captionsetup{width=.9\linewidth}
	  \includegraphics[width=\linewidth]{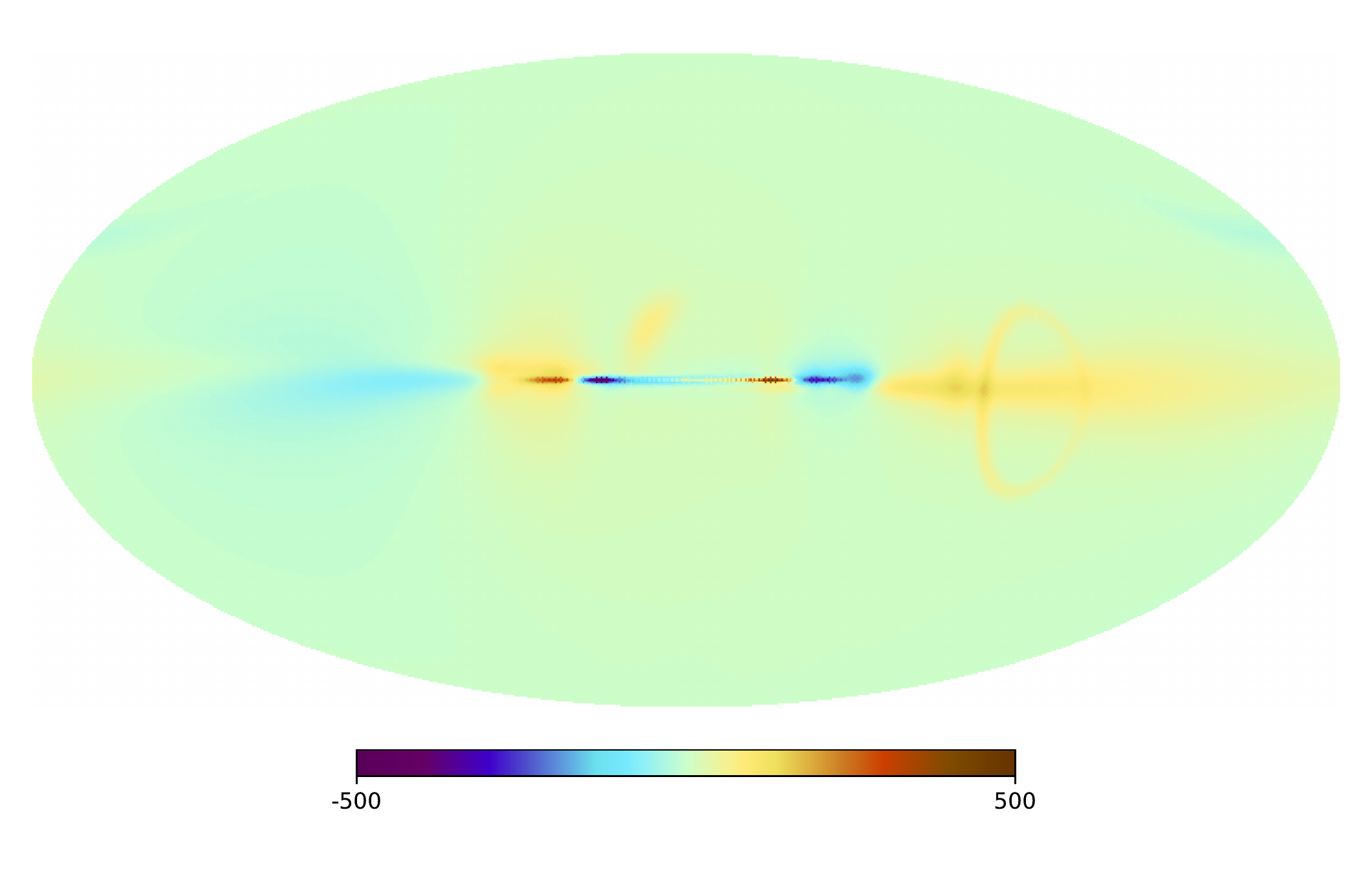}
  \caption{Map of the Galactic Faraday depth produced with \hammurabiX\ based on the WMAP LSA model in $\mathrm{rad/m^2}$. The model's parameters were set to $B_0=-0.51~\mathrm{\mu G}$, $\psi_0=280^\circ$, $\psi_1=330^\circ$, and $\chi_0=30.3^\circ$.}
\end{subfigure}
\label{fig:Faraday_depth}
\end{figure*}

\begin{figure*}
\caption{Streamplot of the WMAP LSA model for $B_0=-0.51 ~\mathrm{\mu G}$, $\psi_0=280^\circ$, $\psi_1=330^\circ$, and $\chi_0=30.3^\circ$.}
  \centering
  	\includegraphics[width=.9\linewidth]{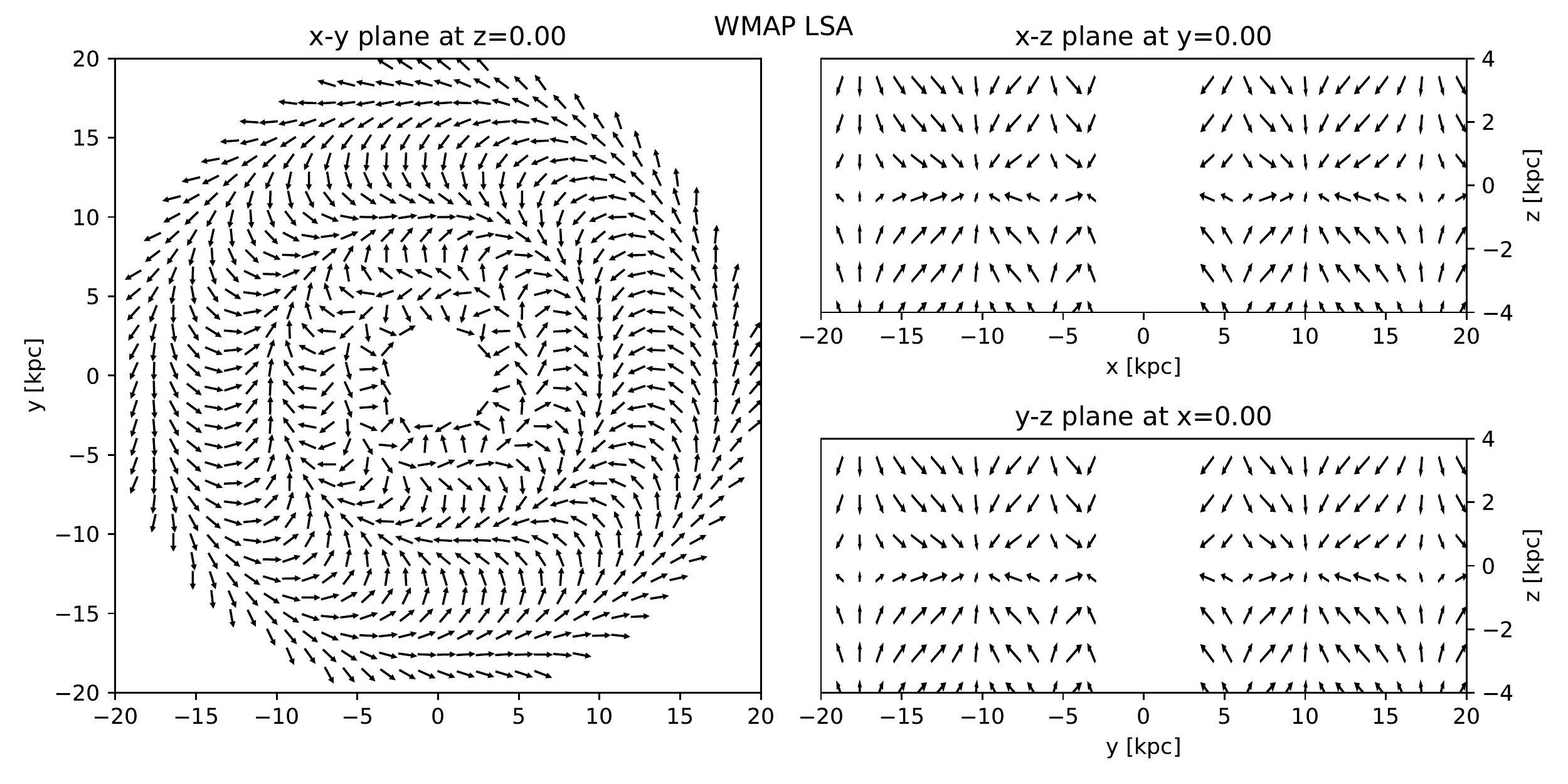}
\label{fig:wmap_field_lines}
\end{figure*}

When using Faraday depth instead of synchrotron data \citep{2012A&A...542A..93O} for a parameter fit, cf.\ \ref{fig:marginals_data_norandom_fd}, the limited capabilities of the WMAP LSA model become clear. 
Even though the WMAP LSA model was designed for fitting synchrotron radiation but not Faraday depth data, it is nevertheless remarkable how incompatible they are. 
Not only are the estimates for $\psi_0=280^\circ$ and $\psi_1=330^\circ$ far off their reference values, the magnetic field strength is pushed to values near zero.
The latter indicates a general incompatibility between the model and the data.  
Furthermore, the best fit value for $B_0$ is negative, which in our case means that the direction of the magnetic field is reversed compared to \citet{page:2007}.  
\ref{fig:Faraday_depth} illustrates the issue. 
In the data, one can locate a dipole as well as a quadrupole moment both being aligned with the Galactic plane. 
Because of its simple structure, the WMAP LSA model cannot account for the double anti-axisymmetric quadrupole structure.
That is expected, and if such a feature is needed one can use more complex models like JF12 \citep{jansson:2012c} or Jaffe13 \citep{jaffe:2013}. 
But, beyond this, \ref{fig:Faraday_depth} in combination with \ref{fig:wmap_field_lines} reveals that the likelihood peak at $\psi_1 = 330.2^\circ$ corresponds to a configuration where the model exhibits field reversals to fit the structure in the Galactic plane. 
Although the Faraday rotation map compares well to the data, such a parameter configuration is the result of a simple model fitted to a complicated dataset and is not necessarily the most physically realistic solution.
This demonstrates another possible pitfall and also how important it is to incorporate physical priors for the model parameters when doing a real-life analysis. 
\imagine\ provides the structure for comprehensive studies that are not only built on powerful algorithms such as \multinest\ that will find parameter estimates in any case but also regularizes them and points out problems in the reconstruction. 
Furthermore, irrespective of the quadrupole, for the reference parameter values also the dipole does not fit; it has the wrong sign. 
This means that the overall field orientation itself in the WMAP LSA model cannot be correct. 
This is a fact that does not become apparent when solely using synchrotron data, since even though synchrotron emission is sensitive to the magnetic field's direction, it is not to its orientation.
Using the \imagine\ pipeline for parameter estimation, it is economic to include various data sets from different observables, since the \imagine\ data repository is open and will grow through collaborative contribution. 
Such obvious contradictions can then be avoided by a more comprehensive approach. 

\begin{figure*}
\caption{\textbf{Synchrotron radiation data, WMAP LSA plus random magnetic field} Marginalized posterior plots and projected pairwise correlation plots from applying \pymultinest\ to $408~\mathrm{MHz}$ and $30~\mathrm{GHz}$ synchrotron data from \citet{2016A&A...594A...1P}. 
The dashed lines represent the $16\%$, $50\%$ and $84\%$ quantiles, respectively. 
}
\centerline{\includegraphics[width=1.15\linewidth]{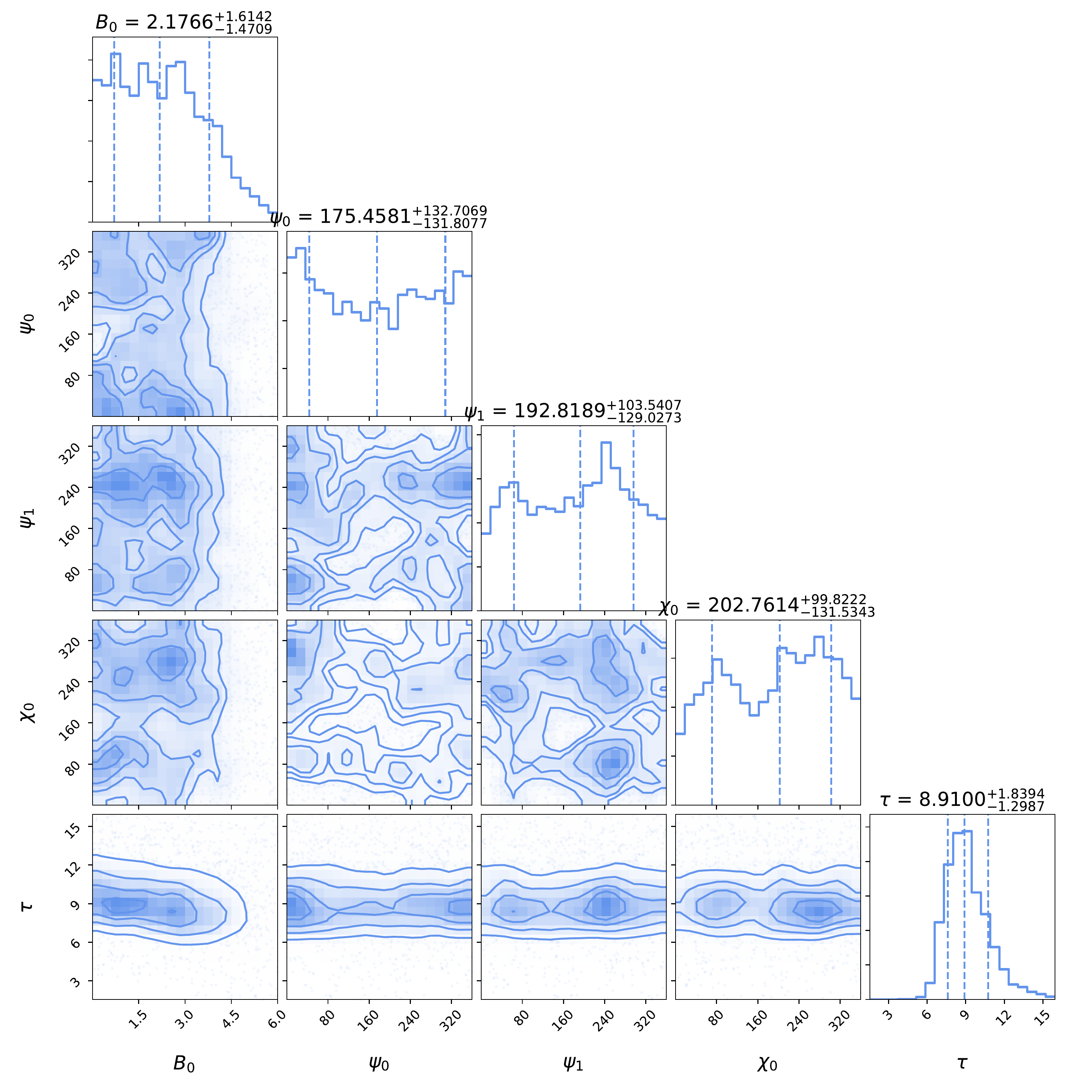}}
\label{fig:marginals_data_random_sync}
\end{figure*}

Going a step further, we try to find parameter estimates for the WMAP LSA plus random magnetic field model that we previously used for the mock data tests in \ref{sec:mock_data_tests}.
The results are shown in \ref{fig:marginals_data_random_sync}. 
With respect to the random magnetic field, we limit ourselves to the inference of $\tau$, the strength of the random magnetic field. 
For the sampling we used a wide prior volume, especially for the angular parameters $\psi_0$, $\psi_1$ and $\chi_0 \in \left[0, 360\right]^\circ$, each.
Note that only $\psi_0$ is a truly circular parameter, cf.\ \ref{eq:WMAP}, and thus was setup as such in \pymultinest.  
Comparing the results of a fit based purely on synchrotron emission, given in \ref{fig:marginals_data_random_sync}, to the scenario in \ref{fig:marginals_data_norandom_sync} with only an ordered field provides several insights.
First of all, we see how approaches that neglect the Galactic variance tremendously underestimate uncertainties when doing parameter estimation.
The estimate for $B_0$ is smeared out the least, but the predictive power for $\psi_0$, $\psi_1$ and $\chi_0$ disappears when taking the Galactic variance into account correctly.
In the light of the above, it is noteworthy how clear the prediction for the strength of the random magnetic field turns out to be.
All in all, despite their weak predictive power, the results shown in \ref{fig:marginals_data_random_sync} are consistent with those in \ref{fig:marginals_data_norandom_sync}.
For the former, the regular magnetic field strength $B_0$ is smaller compared to the latter as now the random magnetic field also contains magnetic field power in $\tau$.
Also note the reasonable anti-correlation between $B_0$ and $\tau$.  
The overall order of magnitude of $\tau$ is compatible with \citet{2006ChJAS...6b.211H}.
Since $\psi_0$ is a circular parameter it is necessary to consider circular definitions of mean and standard-deviation \citep{watson1983statistics}, which yield $\psi_0 = 0.66 \pm 2.60 ^\circ$. 
Hence, the estimate for $\psi_0$ points towards the same order of magnitude as $\psi_0 = 6.69^\circ$ shown in \ref{fig:marginals_data_norandom_sync}. 
Furthermore, for $\chi_0$ we see a peak around $80^\circ$ which can be interpreted to correspond to the previous best fit value $\chi_0 = 78.6^\circ$. 
The second peak around $\chi_0 = 260^\circ$ is less clear and is likely to be a morphological degeneracy.
As synchrotron emission is sensitive to the magnetic field's direction but not to its orientation, considering \ref{eq:WMAP}, we expect a diffuse degeneracy in $\chi_0$, which gets disturbed by the factor $\tanh\left(\nicefrac{z}{z_0}\right)$.

\begin{figure*}
\caption{\textbf{Faraday rotation data, WMAP LSA plus random magnetic field} Marginalized posterior plots and projected pairwise correlation plots from applying \pymultinest\ to Faraday rotation data from \citet{2012A&A...542A..93O}. 
The dashed lines represent the $16\%$, $50\%$ and $84\%$ quantiles, respectively. 
}
\centerline{\includegraphics[width=1.15\linewidth]{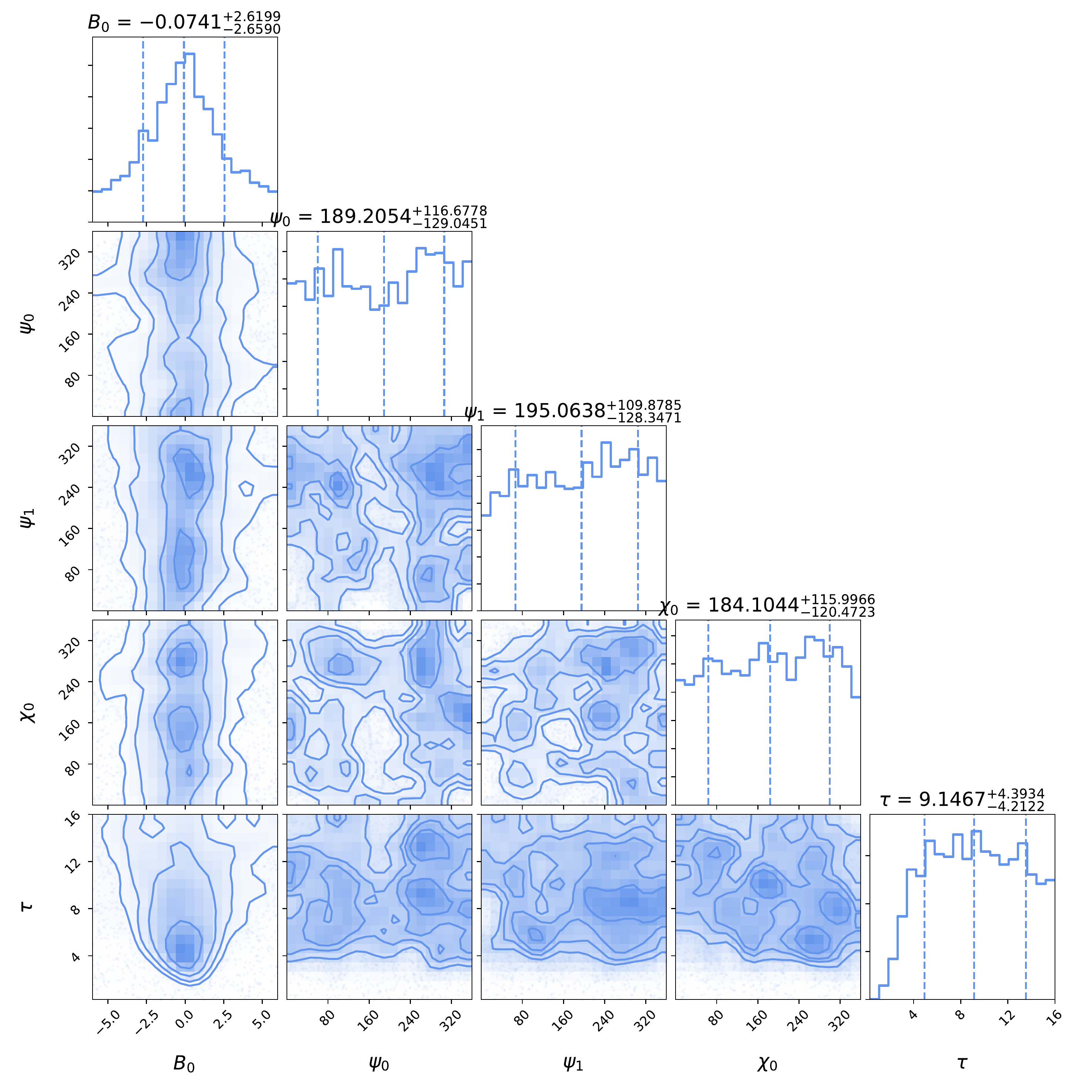}}
\label{fig:marginals_data_random_fd}
\end{figure*}

Repeating this analysis for Faraday rotation data from \citet{2012A&A...542A..93O}, results shown in \ref{fig:marginals_data_random_fd}, underlines what has been seen in \ref{fig:marginals_data_norandom_fd}. 
The WMAP LSA model is inherently incompatible to Faraday rotation observations: $B_0$ is pushed to values near zero and an increasing $\tau$ solely broadens $B_0$'s likelihood as discussed in \ref{sec:likelihood} but does not add anything to the intrinsic quality of fit. 
The likelihoods for $\psi_0$, $\psi_1$ and $\chi_0$ don't posses any clear peaks nor pairwise correlations. 

All in all, this simple example of the WMAP LSA model augmented with a random magnetic field already illustrates how challenging it is to create models of the constituents of the Galaxy with consistent geometry and to find reliable estimates for their parameters. 
While the example in this section was rather academic due to the model's simplicity, the presented steps similarly apply when analyzing more complex models such as JF12 and Jaffe13.
		
\section{Conclusion \& Outlook}
In this paper we presented \imagine, a framework for GMF model parameter inference. 
We have discussed the motivation behind Bayesian parameter inference and model comparison as well as the importance of the \emph{Galactic variance}.
We then described the modular structure and extensibility of the \imagine\ framework. 
Its most important building blocks are: 
\begin{itemize}
\item state-of-the-art parametric GMF models, 
\item a varied set of complementary observables,
\item the new and improved \hammurabiX\ simulator, and 
\item the different sampling algorithms that can be used within \imagine. 
\end{itemize}
In \ref{sec:application}, we showed with mock data that the pipeline works self-consistently, 
	we illustrated the concept of Bayesian model comparison, 
	and we then applied the pipeline to real data. 
In the course of this, we showed the importance of multi-observable based parameter fitting.
This analysis was, however, a simple proof-of-concept to demonstrate the capabilities of the \imagine\ pipeline. 
Now, more sophisticated analyses are in order to gain as much scientific insight from existing data sets and GMF models as possible. 
Since \imagine\ is uniquely suited to handle the random component of the GMF and its uncertainties correctly, it can be adapted into a powerful tool to study the turbulent ISM by, e.g., adding a structure function analysis to the likelihood in order to constrain the turbulent spectral index. 
All those insights should be used to build improved models and to keep the models' best-fit parameter estimates up-to-date with respect to the ever improving data. 
Within this paper we solely inferred the parameters of the GMF while keeping the thermal electron density fixed. 
With an extended list of observables (e.g., the DM), more informative datasets, and better models, we can extend this work to a joint inference of the magnetic field, the thermal electron density, the cosmic ray population, and even the dust model parameters.  
The \imagine\ pipeline is ready to help tackle this challenge and is available at: \href{https://gitlab.mpcdf.mpg.de/ift/IMAGINE}{https://gitlab.mpcdf.mpg.de/ift/IMAGINE}

\begin{acknowledgements}
	  We thank François Boulanger, Martin Reinecke, Luiz F. S. Rodrigues, and Anvar Shukurov for fruitful discussions and valuable suggestions. 
      Part of this work was supported by the \emph{Studienstiftung des deutschen Volkes}.
      The original concept of \imagine\ arose from two International Team meetings\footnote{\href{http://www.issibern.ch/teams/bayesianmodel/}{http://www.issibern.ch/teams/bayesianmodel/}} hosted by the International Space Science Institute in Bern. 
      We also acknowledge support and hospitality of the Lorentz Center in Leiden, where the \imagine\ project was further discussed and refined.\footnote{\href{http://www.lorentzcenter.nl/lc/web/2017/880/info.php3?wsid=880}{http://www.lorentzcenter.nl/lc/web/2017/880/info.php3?wsid=880}} 
	  We acknowledge the support by the DFG Cluster of Excellence ”Origin and Structure of the Universe”.
	  The computations have been carried out on the computing facilities of the Computational Center for Particle and Astrophysics (C2PAP) and the Radboud University, Nijmegen, respectively. 
      This research has been partly supported by the DFG Research Unit 1254 and has made use of the NASA/IPAC Infrared Science Archive, which is operated by the Jet Propulsion Laboratory, California Institute of Technology, under contract with the National Aeronautics and Space Administration.
      Some of the results in this paper have been derived using the \healpix\ \cite{2005ApJ...622..759G} package. 
      The corner plots where made using the \texttt{corner} \python\ package \citep{corner}. 
\end{acknowledgements}

\bibliographystyle{aa}
\bibliography{imagine}

\end{document}